\documentclass[fleqn,usenatbib]{mnras}

\usepackage{newtxtext,newtxmath}

\usepackage[T1]{fontenc}

\DeclareRobustCommand{\VAN}[3]{#2}
\let\VANthebibliography\thebibliography
\def\thebibliography{\DeclareRobustCommand{\VAN}[3]{##3}\VANthebibliography}

\usepackage{tabularx,booktabs,multirow,makecell}
\usepackage{graphicx,array}	

\graphicspath{{./}{./images/}}
\DeclareGraphicsExtensions{.pdf,.png,.jpg}

\usepackage{enumitem}
\setlist[enumerate]{topsep=0pt,wide,labelwidth=!,itemindent=!, leftmargin=0em} 

\usepackage{bm}
\usepackage{amsmath}	
\usepackage[dvipsnames]{xcolor}

\newcommand{\Hi}{H\textsc{i}}

\newcommand{\snia}{SN\textsc{i}a}
\newcommand{\snii}{SN\textsc{ii}}
\newcommand{\lr}{\gamma}

\newcommand{\ie}{i.e.}

\newcommand{\N}{$\mathcal{D}$}
\newcommand{\Nr}{$\mathcal{D}_r$}

\newcommand{\sven}[1]{{\color{Black}{#1}}}
\definecolor{dogwoodrose}{rgb}{0.84, 0.09, 0.41}

\newcommand{\mc}[1]{#1}

\sven{\title[A study of bubbles in a simulation]
{More than a void? The detection and characterization of cavities 
in a simulated galaxy's interstellar medium}}

\author[A. Taghribi et al.]{Abolfazl Taghribi,$^{1}$\thanks{E-mail: abolfazl.taghribi@gmail.com}
Marco Canducci$^{4}$,
Michele Mastropietro$^{3}$,
Sven De Rijcke$^{3}$,
Reynier Frans Peletier$^{2}$,\newauthor
Peter Tino$^{4}$,
and Kerstin Bunte$^{1}$
\\
$^{1}$University of Groningen, Bernloulli Institute of Mathematics, Computer Science and Artificial Intelligence, 
P.O. Box 407 9700 AK Groningen, NL\\
$^{2}$University of Groningen, Kapteyn Institute, P.O. Box 800 9700 AV Groningen, NL\\
$^{3}$Ghent University, Department of Physics \& Astronomy, Krijgslaan 281, S9, B-9000 Ghent, Belgium\\
$^{4}$University of Birmingham, school of computer science, B15 2TT, Birmingham, United Kingdom
}

\date{Accepted 18 Dec. 2024; Received 22 Apr. 2024; Available online 1 Jan. 2025}

\pubyear{2025}

\begin{document}
\label{firstpage}
\pagerange{\pageref{firstpage}--\pageref{lastpage}}
\maketitle

\begin{abstract}
The interstellar medium of galaxies is filled with holes, bubbles, and shells, typically interpreted as remnants of stellar evolution. There is growing interest in the study of their properties to investigate stellar and supernova feedback. So far, the detection of cavities in observational and numerical data is mostly done visually and, hence, is prone to biases. Therefore, we present an automated, objective method for discovering cavities in particle simulations, with demonstrations using hydrodynamical simulations of a dwarf galaxy. The suggested technique extracts holes based on the persistent homology of particle positions and identifies tight boundary points around each.
With a synthetic ground-truth analysis, we investigate the relationship between data density and the detection radius, demonstrating that higher data density also allows for the robust detection of smaller cavities. By tracking the boundary points, we can measure the shape and physical properties of the cavity, such as its temperature.
In this contribution, we detect 808 holes in 21 simulation snapshots. We classified the holes into supernova-blown bubbles and cavities unrelated to stellar feedback activity based on their temperature profile and expansion behaviour during the 100 million years covered by the simulation snapshots analysed for this work. Surprisingly, less than 40\% of the detected cavities can unequivocally be linked to stellar evolution. Moreover, about 36\% of the cavities are contracting, while 59\% are expanding. The rest do not change for a few million years. Clearly, it is erroneous to interpret observational data based on the premise that all cavities are supernova-related and expanding.
This study reveals that supernova-driven bubbles typically exhibit smaller diameters, larger expansion velocities, and lower kinetic ages (with a maximum of 220 million years) compared to other cavities.
\end{abstract}




\section{Introduction}

It has been appreciated since the late 1960s that the interstellar medium of external galaxies is filled with holes, bubbles, and shells with a wide range of radii and sizes \citep{1966MNRAS.131..371W, 1967AuJPh..20..147H, 1967PASP...79...29H}. Such structures have been observed in both optical and radio data. A decade later, radio surveys of the Milky Way revealed similar structures in our own galaxy \citep{1976ApJ...208L.137H}. Since then, observations have progressed beyond detections and now include detailed quantitative studies of these bubbles, characterizing their size distributions, spatial distributions, and kinematics \citep{1997MNRAS.289..570O, Bagetakos2011, Nath2020, pokhrel_catalog_2020}. For the dwarf galaxies near the Milky Way, exceptionally high spatial resolution data is now available. For instance, \citep{2003ApJS..148..473K} merged ACTA and Parkes radio observations of the Large Magellanic Cloud and succeeded in measuring the properties of {\Hi} cavities on scales from 15 parsecs upwards.

Understanding the properties and origin of these bubbles or cavities is an important astrophysical problem because of the possible link between such bubbles and stellar evolution. The idea that stellar winds from massive stars could be responsible for these bubbles was first introduced by \citet{1975ApJ...200L.107C}. If this is the case, then these bubbles are intimately linked to the subgrid physics employed in numerical simulations of (dwarf) galaxy formation and evolution \citep{2011MNRAS.416..601S}. Thus, bubble properties offer a testbed for the subgrid recipes for stellar and supernova feedback, which regulate the energy injection into the interstellar medium and are expected to affect the {\Hi} structure. As such, comparing the properties of the {\Hi} cavities in real and simulated dwarf galaxies is expected to help constrain the physics of galaxy evolution.

If such a one-to-one relation between newborn stars or star clusters, on the one hand, and {\Hi} cavities on the other exists, then one would expect some observational tracer of recent or ongoing star formation to correlate with the {\Hi} structure. However, H$\alpha$ and 24-micron emission, which trace ongoing star formation, appear to correlate poorly with {\Hi} cavities \citep{2008ApJS..175..165B}. Moreover, detailed observations of the dwarf galaxy Holmberg~{\sc ii} with the Hubble Space Telescope showed that the stars inside a given {\Hi} cavity are far from coeval, and their ages do not reflect the kinematical age of the host cavity \citep{2009ApJ...704.1538W}. Instead, these authors propose a multi-age model, in which stars born over several tens of megayears contribute to inflating an {\Hi} cavity. The observed correlation between UV emission, which traces young but not necessarily newborn stars, and {\Hi} structure appears to corroborate this idea. Recent JWST observations of nearby galaxies suggest that their intricate gas structures may actually be unrelated to stellar feedback \citep{2023ApJ...944L..18M}. The cavities visible in the interstellar medium of these galaxies have dimensions comparable with the local turbulent Jeans length. This was interpreted as evidence for three-dimensional fragmentation of the gas disk as the root cause of the observed network of filaments and voids. This fragmentation is expected to be regulated by the balance between turbulent pressure and self-gravity, leading to a correlation between the structures' sizes and the Jeans scale.

Cavities are often identified via a visual inspection of the observed {\Hi} column density distribution \citep{2003ApJS..148..473K} or CO emission maps \citep{2023A&A...676A..67W}, possibly extended with a visual inspection of the velocity channel maps, as in \citet{pokhrel_catalog_2020}. Sometimes, as in \citet{2023A&A...676A..67W}, a correlation between the gas structures and the stellar-population properties is required for cavity detection. Observationally, only projected position and line-of-sight velocity information are available, imposing limitations on data interpretation. For instance, an expanding shell, with its front side approaching the observer and its back side receding, looks identical to an imploding shell, with its back side approaching the observer and its front side receding. Interpreting the data based on the assumption that all shells are expanding may lead to false conclusions. This reliance on human judgment introduces a degree of subjectivity and a potential lack of reproducibility in the bubble identification process, and a more automated procedure is clearly desirable.

Systematic studies and comparisons of existing void-finding algorithms to detect two- and three-dimensional cosmic voids in the context of the cosmic web and Cosmic Microwave Background (CMB) are provided by \cite{colberg_aspen--amsterdam_2008,cautun_santiagoharvardedinburghdurham_2018,Feldbrugge_2019}. 
To the best of our knowledge, the community has not yet converged to a clear definition of the spatial boundary of these generic low-density regions, and hence many different ways were proposed to identify them. Strategies include identification based on spherical underdensity, space tessellation, and variations thereof, which will influence the shape of the resulting voids and potential overlap. They generally only partially overlap in the data they use, and some require modalities generally not available in observations. The aforementioned studies 
did not explicitly aim to find the ``best'' void-finder algorithm but instead, provide a good overview of existing identification strategies and investigate where they agree and where not. They argue about situational beneficial choices with respect to cosmic-void data from simulation and observation.

Topological data analysis, or TDA, provides exploratory tools to examine $N$-dimensional data, and it presents general information about the topology of data such as shape, connectivity, and holes in any dimensions \citep{edelsbrunner_shape_1983,edelsbrunner_three-dimensional_1994, Edelsbrunner2002, CARLSSON2005, boissonnat_geometric_2018, edelsbrunner2022computational}. 
TDA tools 
have extensively been deployed in astronomy and astrophysics 
for the identification of filaments, walls, and voids in the cosmic web \citep{platen_cosmic_2007, van_de_weygaert_cosmic_2009, van_de_weygaert_alpha_2011,sousbie_persistent_2011-1, sousbie_persistent_2011, Pranav2016}, as well as the comparison of astrophysical simulations of the interstellar medium of galaxies with observations \citep{makarenko_topological_2018}.
Most of the methods exhibit a primary focus on the distribution of Betti numbers, which quantify the proportions of clusters, filaments, walls, and voids. Zobov \citep{Neyrinck2008} and its extended toolkit VIDE \citep{SUTTER2015} employs Voronoi tessellation for density estimation to identify voids. 
However, TDA is sensitive to limitations and changes in data samples, such that the structure and shape of cosmic web components will strongly depend upon the sample used. 
This existence of topological bias caused by sampling effects is a persistent problem for the analysis of structures and might have adverse effects on physical conclusions drawn as indicated in \citep{Bermejo2024}.

The work of \citep{XU2019} identifies voids and filaments in the cosmic web, that constitutes the study most closely related to our work. 
In contrast to the analysis of large-scale-structure (LSS) our research focuses on smaller-scale structures, specifically dwarf galaxy simulations, where the properties of individual topological features are of particular interest. 
The recovered boundaries suggested by \citep{XU2019} are often not precisely aligned with the edges of voids or cavities, showing variability with repeated sampling of the point cloud. 
They also construct the filtration on a 3D grid and compute the distance-to-measure function for each grid point, leading to boundary point fluctuations when the grid size changes. 
Moreover, a fine grid resolution for accurate void structure recovery significantly increases computational cost.
Generally the computational complexity of TDA tools poses a challenge, restricting 
their application to very large and higher dimensional datasets.
Therefore, in practice one limits the dimension of the features considered and typically approximates specific filtrations, see \cite{otter_roadmap_2017}. 
Despite these improvements the computation is still very challenging and poses a hard limit on the filtrations on large data sets \citep{Stolz_2023}.

An alternative to filtration approximations is the subsampling of the data with selected landmarks as a reduced representation of big data sets. Mainly two standard approaches to select landmarks exist, namely \emph{uniform random selection} and selection via the MaxMin algorithm \citep{de2004topological}.  Neither is ideal and in particular, the MaxMin algorithm tends to include outliers \citep{Stolz_2023}. 
The latter therefore proposed a K-means based landmark selection optimized for persistent homology, that is robust against outliers. 
However, their algorithm requires large sampling densities and exhibits high computational costs, and hence still limits the applicability in large astronomical data sets.
Recently our proposed \textsc{asap} (Sub-sampling Approach for Preserving topological structures, \citet{taghribi_asap_2021}) introduces a subsampling technique to reduce the computational complexity. The authors demonstrate its effectiveness in mitigating computational demands. Furthermore, that contribution presents a robust method for extracting tight boundaries around holes and provides a probabilistic model for their irregular shapes. The paper thus equips researchers with practical tools for studying bubbles within large-scale datasets, overcoming previous limitations in size and required resources.

In this paper, we analyse bubbles within a simulation of a jellyfish-like dwarf galaxy, contributing in four aspects:
1) We demonstrate the effectiveness of \textsc{asap} \citep{taghribi_asap_2021} in combination with TDA for detecting gas particles located on the boundaries of bubbles in particle simulations.
2) Following the identified hull particles before and after the cavity's discovery, we explore various physical properties.
3) We compute characteristics known from observational studies \cite{1997MNRAS.289..570O, pokhrel_catalog_2020} and compare them.
4) Finally, a controlled empirical study with synthetic data investigates the robust detection and analysis of bubbles in datasets with particle densities corresponding to different regions of the jellyfish simulation.
Studying bubbles and holes in the simulation and comparing their properties with those of observed galaxies provides concrete evidence about the processes that create these structures, their transformation over time, and their impact on galaxies.

This paper is structured as follows: \autoref{sec:simulation} presents information about the $N$-body/SPH simulation of a dwarf galaxy, and \autoref{sec:holedetection} details the pipeline and methods for extracting cavities and particles on their boundaries. Subsequently, \autoref{sec:holeanalysis} examines a range of properties of several example bubbles and investigates multiple characteristics of all detected cavities in the particle simulation snapshots under study. We conclude in \autoref{sec:conclusion}.

\vspace{-0.3cm}

\section{Simulation data}\label{sec:simulation}

We use the MoRIA (Models of Realistic Dwarfs In Action) suite of simulations \citep{Verbeke2017}. 
This is a set of high-resolution $N$-body/SPH simulations of late-type dwarf galaxies, evolved both in isolation and in a Fornax Cluster-like environment \citep{2015ApJ...815...85V, Mastropietro2021}.  
The simulation snapshots used in this work (available at \url{https://github.com/abst0603/ASAP}) represent the gas and stars present in a galaxy as a cloud of points. 
Each particle carries information about its position and velocity, as well as its chemical composition and, in the case of a gas particle, its density and temperature. 
For the present study, we use a simulation of a dwarf galaxy evolving inside a cluster environment because this allows us to explore the detection of bubbles inside environments covering a wide variety of densities.
All numerical and physical details of 
these simulations can be found in the referenced papers. 

\vspace{-0.2cm}

\subsection{Feedback parameters}

One expects at least a subset of all bubbles detected in a galaxy to have been caused by events that inject energy into the interstellar medium, such as stellar supernova explosions. Therefore, we clarify the energy feedback recipes employed in the simulations here.

Each stellar particle in a simulation represents a single stellar population (SSP): a group of stars born from the same gas cloud with identical characteristics (\ie{} age and metallicity).
In these simulations, a gas particle becomes eligible for conversion into a star particle, thus mimicking star formation, when the gas density exceeds 100~amu~cm$^{-3}$.
At birth, the stellar masses follow a Chabrier initial-mass function \citep{2003ApJ...586L.133C}.
At the end of their short lifetimes (between $4$ and $31$ Myr) on the main sequence, stars with masses in the interval $8-70$~M$_\odot$ explode as type II supernovae (\snii), each event injecting $10^{51}$~ergs of energy into the surrounding interstellar medium.
This energy is returned continuously by a stellar particle over the $4-31$~Myr age interval.
Less massive stars in binary systems can lead to type Ia supernovae (\snia), which also produce $10^{51}$~ergs per event. 
The {\snia} delay times are normally distributed around a mean delay of $~4$~Gyr with a $0.8$~Gyr dispersion \citep{2004ApJ...613..200S}.

To limit the computational overhead caused by returning very small amounts of {\snia} feedback originating from the tails of the time-delay distribution, we limit feedback to a $3\sigma$ time interval around the mean. 
The ratio of the number of {\snia} to {\snii} is set to 0.15.
Stellar winds, or SW, release $10^{50}$~erg per stellar particle weighted with the fraction of the  particle's mass in the form of stars in the $8-70$~M$_\odot$ mass range. 
SW energy is returned from the moment a stellar particle is born until the least massive star goes supernova.
Besides energy, {\snii} also return 19~\% of the mass of the stellar particle to the surrounding gas particles.
{\snia} and SW return a mere 0.6~\% of the particle's original mass. 
Stellar particles with a metallicity below $\text{[Fe/H]}=-5$ are treated as Pop~III particles.
As detailed in \citep{2015ApJ...815...85V}, Pop~III particles return 4 times as much energy via {\snii} and 40 times as much energy via SW as a ``normal" stellar particle \citep{2010ApJ...724..341H}.
Pop~III particles lose 45~\% of their original mass to the surrounding gas.

For the remainder of the study, we applied a threshold on the density of gas particles and saved the points with a density value $\rho \geq 10^{-3}~ \mathrm{atom/cm^3}$. 
This limit in density is of the order of magnitude of the self-shielding threshold implemented in the code \citep[eq.~(5) in][]{DeRijcke2013}: 
if the gas is dense enough to be able to self-shield, its hydrogen can emit in the 21cm line, the most used band to observe the bubbles under study.
In other words, the density threshold is physically motivated to select particles emitting in \Hi{}.
Also, in such low density regions, the distance between particles is larger than holes inside the galaxy, and this sparsity introduces numerous spurious holes to the study.
Full details on the simulation can be found in \citet{Mastropietro2021}.




\section{Method for automatic bubble detection} \label{sec:holedetection}


A Sub-sampling Approach for Preserving Topological Structures (\textsc{asap}) was proposed in \citet{taghribi_asap_2021} to overcome computational challenges posed by TDA tools. Note that this subsampling can also be used with existing void finders that are not density-based, such as tessellation approaches. With subsequent TDA and a voting strategy, \textsc{asap} was demonstrated to robustly detect particles that compose bubble boundaries. We summarize \textsc{asap} and subsequent efficient bubble detection in the context of a hydrodynamical galaxy-evolution simulation below and refer the interested reader to \citet{taghribi_asap_2021} for full details. The implementation is available at: \url{https://github.com/abst0603/ASAP}.

\begin{figure}
\centering
    \includegraphics[width=\columnwidth]{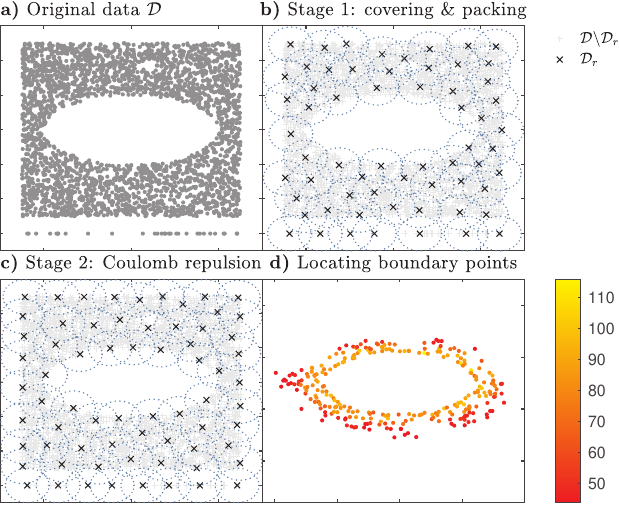}
    \caption{
    \textsc{asap} and voting for cavity boundary detection shown on a holed square with 2970 and line with 30 points. 
    Stage 1 shows the resulting 68 subset \mc{\Nr} ensuring covering and packing of the original data \mc{\N} using radius $r=0.5$. 
    Stage 2 applies Coulomb-like repulsion to achieve a more uniform distribution of 
    \mc{\Nr}. d) shows the votes for boundary points (see section \ref{sec:holelocation}).}
    \label{fig:asap}
\end{figure}

Assume the available dataset {\N} consists of the three-dimensional position vectors of $N$ particles. 
Then \textsc{asap} aims to find a subset \Nr$\subset${\N} 
with much 
fewer points, that approximates the structure (topology) of the full data set (as measured by persistent homology). In other words, assume a dataset as visualized in \autoref{fig:asap}a) with 99\% of its points distributed on a holed square and below it a line with the remaining 1\%. While simple random sampling is very likely to lose the line and deform the hole, \textsc{asap} maintains \mc{both} 
structures. This is achieved in two stages: 
1) an iterative process in which a random vector $\vec{s}$ is taken from \N, inserted into \Nr, and all points from {\N} that are within a ball of radius $r$ centred around $\vec{s}$ are removed. 
This process is repeated until {\N} is depleted. As a result of this process, the final samples encompass all parts of the data, with minimal overlap in their coverage areas.
\begin{align}
    &\textrm{covering } &\text{\Nr} = \{\forall \vec{p} \in \text{\N},
\exists\, \vec{s} \in \text{\Nr} \vert\  d(\vec{p},\vec{s})\le r\}\text{ and }\\
    &\textrm{packing }  &d(\vec{s}_i,\vec{s}_j)>r
\quad\forall \vec{s}_i, \vec{s}_j \in \text{\Nr}\text{ with }i\neq j \enspace. 
\end{align}

2) An iterative displacement strategy, inspired by Coulomb's law, is used to distribute elements in \Nr more \mc{regularly}. 
These two stages are exemplified in \autoref{fig:asap}b) and c) for a synthetic dataset consisting of particles sampled from a square with an oval 
hole and a line below it, as shown in panel a). 
Stage 1) guarantees that the original dataset {\N} is covered with the reduced set \Nr, such that each particle from {\Nr} has the closest neighbor from that set no further than $2r$, but not closer than $r$.
Given that this method will later be applied to identify cavities within an astronomical simulation, it is preferable to have more uniformly distributed samples. Furthermore, the cavity sizes correspond to physical measurements, namely the size of detected holes within the dwarf galaxy simulation. Therefore, the samples should be positioned as close as possible to the cavity boundaries.
In stage 2), the particles in \Nr are repelled by neighbouring samples within a radius of $2r$, inspired by Coulomb-like forces between like-charged particles.
If $\mathcal{N}_i$ represents the sample points in $2r$ radius of $\vec{s}_i$ and $\lr$ denotes the learning rate, the displacement is computed as: 
\begin{equation}
\vec{m}_i = \textrm{disp}(\vec{s}_i)
= \sum_{\vec{s}_j\in \mathcal{N}_i}\frac{\vec{s}_j-\vec{s}_i}
{d(\vec{s}_i,\vec{s}_j)
}\cdot\frac{\lr}
{d(\vec{s}_i,\vec{s}_j)^2 
} \quad, 
\label{chargemovement}
\end{equation}
The learning rate $\lr$ gradually decreases, facilitating the algorithm's convergence.
At each iteration, a displaced particle in \Nr is replaced by a original particle closest from {\N} to that displaced position:
\begin{eqnarray}
	\hat{\vec{s}}_i = {\arg\min}_{\vec{p}_j \in \text{\N}}
	\ d(\vec{p}_j,\vec{s}_i + \vec{m}_i
	) 
 \enspace .
    \label{displacement}
\end{eqnarray}
This typically converges in only 10 iterations. 
The resulting more equidistant, topology-preserving subset (see \autoref{fig:asap}c) is beneficial in astronomical applications, where it is crucial to measure the size of circles, cavities, and streams as accurately as possible. The authors demonstrated in \cite{taghribi_asap_2021} that \textsc{ASAP} retains the topological features of a given dataset with fewer samples, regardless of the filtration type used to compute persistent homology. 
In this paper, the remaining experiments are conducted using alpha filtration \citep{edelsbrunner_shape_1983, edelsbrunner_three-dimensional_1994}.

\subsection{Locating bubbles and their boundary particles}
\label{sec:holelocation}

To investigate the structure and origins of bubbles, it is essential to identify particles located on their boundaries building their shells. 
For this purpose, we employ the topological data analysis (TDA) toolbox \textsc{Dionysus} \citep{Morozov_dionysus}, specifically 
an alpha filtration. 
Using this filtration, \textsc{Dionysus} computes the persistent homology and retains information on the generators of topological features, such as holes, at their respective birth times \citep{cohen-steiner_vines_2006}. Subsequently, we extract the vertices associated with each detected cycle.
However, persistent homology alone does not guarantee that the identified boundaries are sufficiently close to the actual shells 
of the holes. For supernova studies, this precision is crucial, as physical characteristics, such as expansion velocity, depend on accurately identifying particles that 
tightly surround the supernova.
To solve this problem, in \cite{taghribi_asap_2021} the authors suggested a voting procedure based on repeated subsampling of the data with \textsc{asap}, which 
enables the selection of tight shell points.
In summary, the multiset $\bar\Gamma$ denotes the set of distinct points $\vec{b}_i$ detected on the boundary of the same cavity at least once in the repeated \textsc{asap}-driven subsampling of the data. 
The multiplicity $m(\vec{b}_i)$ of a point ($\vec{b}_i\in \bar{\Gamma}$) indicates how many times the point is identified as a border point in the repeated subsampling. Now, a vote for every point $v(\vec{b}_i)$ is a counter computed by summing the multiplicity $m(\vec{b}_j)$ of different shell points $\vec{b}_j$ that fall into an open ball with radius $r$ (similar to the radius of \textsc{asap}) centred around $\vec{b}_i$. In other words, the more boundary points fall into an $r$ neighbourhood of a shell point, the higher the vote, and the more likely it is truly part of the border.

\begin{figure}
\centering
\includegraphics[width=1\linewidth, height=0.7\linewidth]{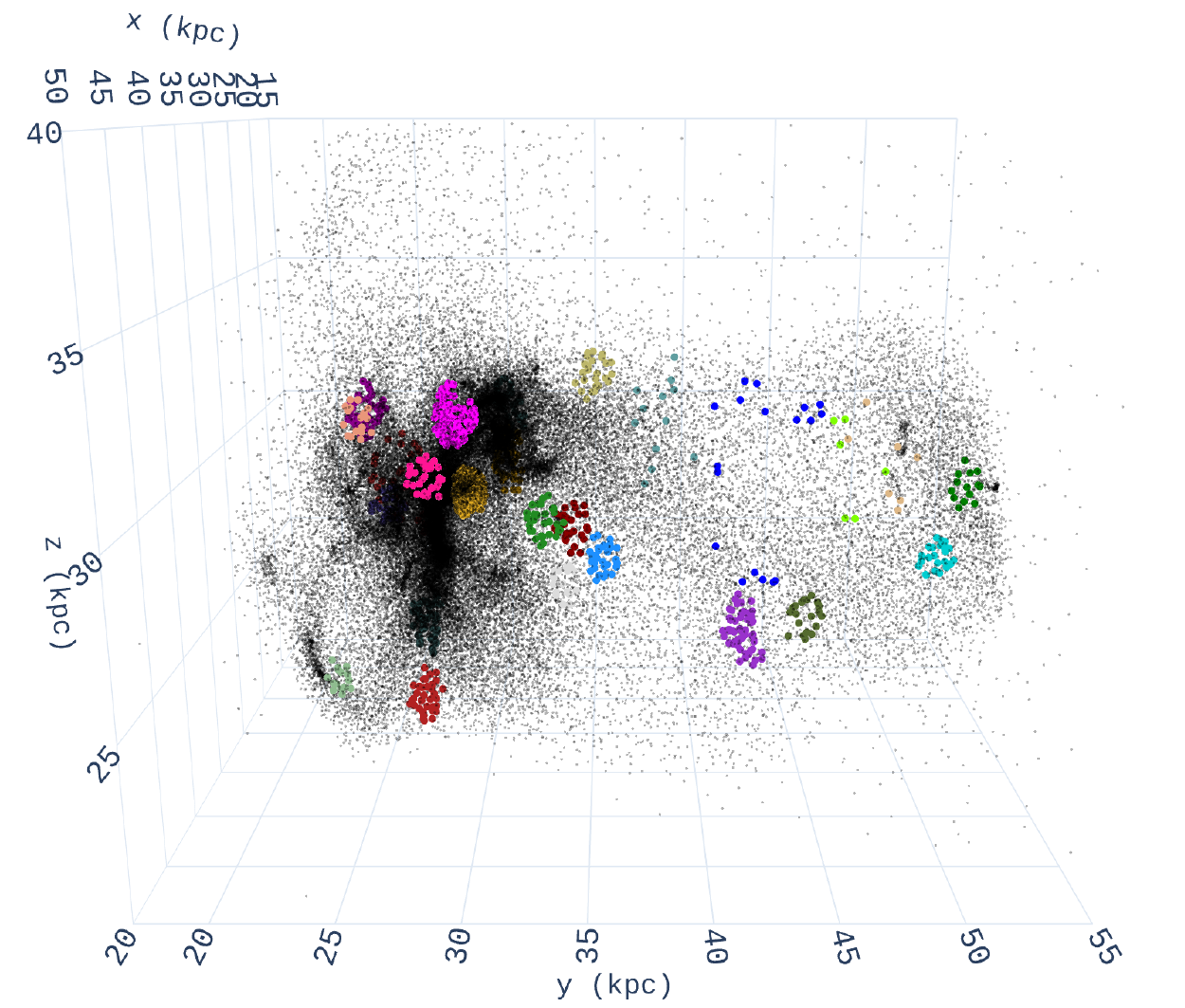}
\\ (a) bubbles (not SN)\\
  \includegraphics[width=1\linewidth, height=0.7\linewidth]{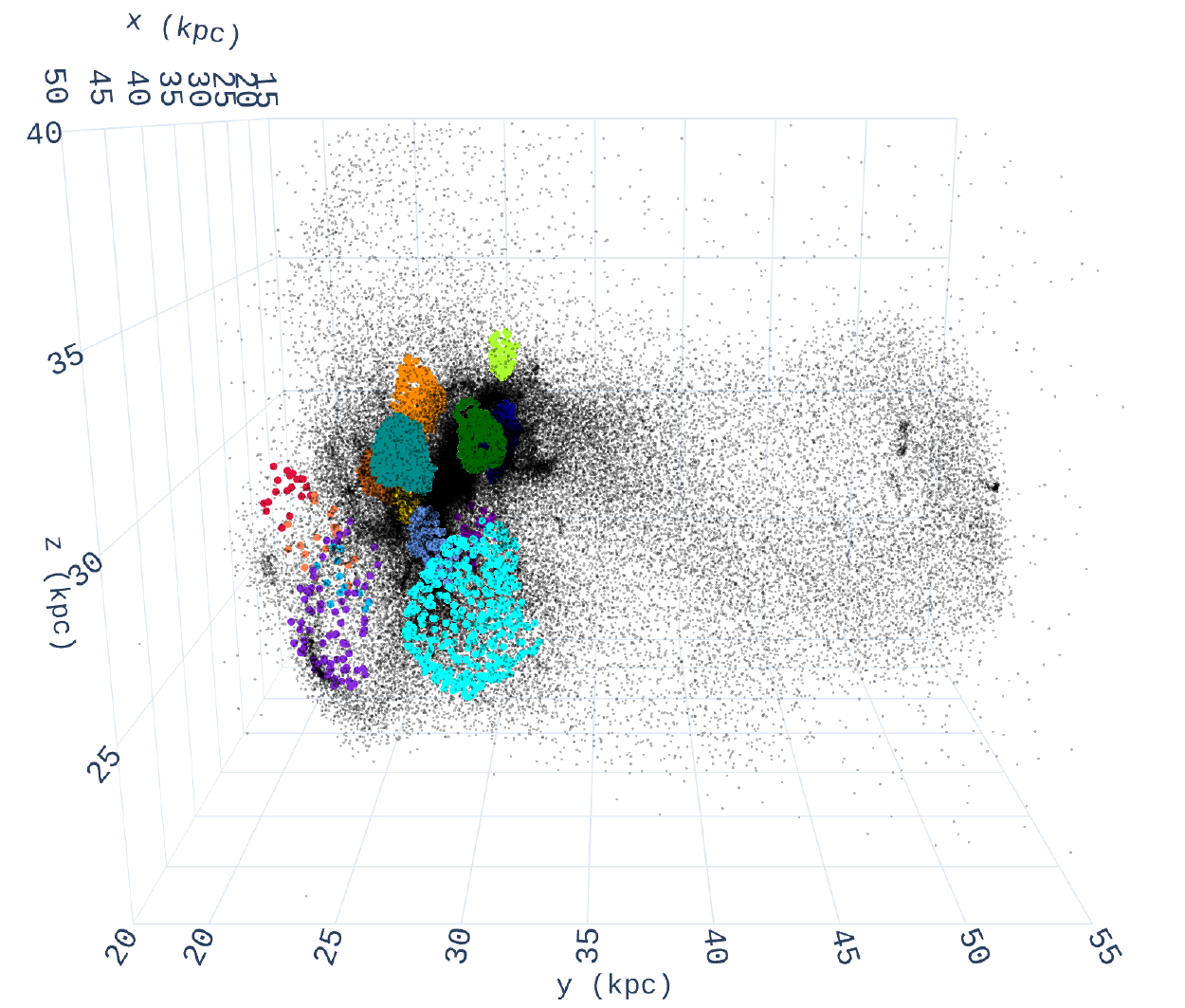}
  \\(b) SN-bubbles
\caption{NonSN-bubbles (a) and SN-bubbles (b) in snapshot 180 of the particle simulation of the jellyfish galaxy.}
\label{fig:BBSN}
\end{figure}

By preserving only the points with a higher number of votes, one can recover a tight boundary around the cavity, as detailed in \citet{taghribi_asap_2021}. 
However, in this study, we save all boundary points since, in particular, small bubbles may only contain very few points on their hull. \autoref{fig:BBSN}(a and b) illustrates all detected holes in a random snapshot (snapshot 180) of the dwarf galaxy simulation, with the particles on their borders highlighted in different colours. 
 
\subsection{The origin of detected bubbles} \label{sec:bubbleorigine}

As observed in \autoref{fig:BBSN}, the detected bubbles in every snapshot differ in size, position inside the simulated galaxy, and the number of particles on their border. In this section, we distinguish between the origin of these holes and group them into two classes, namely bubbles that have a clear link with recent supernova feedback, called SN-bubbles, and those that do not, called nonSN-bubbles.

The jellyfish dwarf galaxy simulation provides a unique identification for every particle through time, and by tracking the detected particles on the border of a hole, one can investigate where the particles were before and how they evolve in the following snapshots. Therefore, for every detected cavity in a snapshot, we track the border particles back and forth through time.

A hole is labelled as a candidate SN-bubble only if both of the following conditions were observed within  
5 snapshots before and after discovering the bubble (11 snapshots in total):
\begin{enumerate}
    \item 
    A surge in the number of particles moving away from the centre of the hole compared to the number of particles moving towards it in one or more  of the snapshots under study.
    \item The mean temperature of the border particles  surpasses 15,000~K in at least one snapshot among the window under study.
\end{enumerate}

\begin{figure}
\centering
\includegraphics[width=1\linewidth]{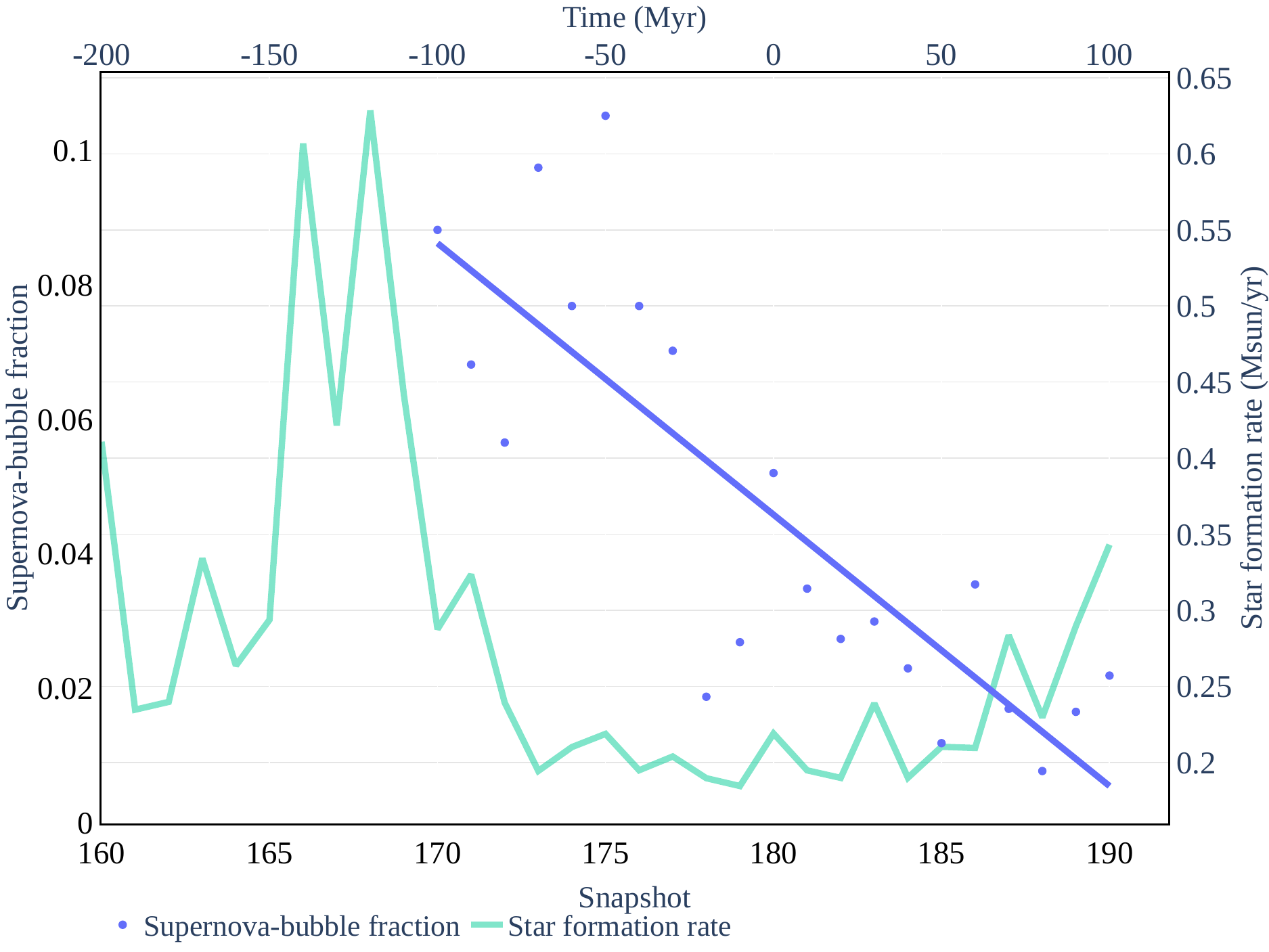}
\caption{Fraction of SN-bubbles, in snapshots 170 to 190 of the simulation introduced in \autoref{sec:simulation} (blue data points), together with their linear regression (blue line).  
\sven{The decline in SN-bubbles follows the star formation rate (teal) with a time delay of roughly 50-75~Myr.}
}
\label{fig:SNpercentage}
\end{figure}

\autoref{fig:BBSN}(b) depicts the coloured boundaries of holes in snapshot 180 of the simulation that fulfil the two conditions mentioned above. We have applied the same pipeline to snapshots 170 to 190 of the same simulation and plotted the proportion of SN-bubbles classified using the two criteria over the range of snapshots in \autoref{fig:SNpercentage}. We observe that within the snapshot window studied, the fraction of SN-bubbles decreases with time.
The decline in the fraction 
of SN-bubbles versus all detected holes in a snapshot follows the additionally plotted star formation rate with a delay.
Nonetheless, there is no clear trend in the number of bubbles formed in every snapshot. 
From snapshot 170 to 190, the number of detected holes ranges from 30 to 47, and the number of SN-bubbles changes from 7 to 26. 
In the remainder of this document, we study different characteristics of the particles on the cavity border to examine whether the hole is likely created by an explosion or not.

\section{Analysis of the detected cavities} 
\label{sec:holeanalysis}

In this section, we examine a variety of characteristics of SN-bubbles and non-SN-bubbles that were detected in snapshots 170 to 190 of the jellyfish galaxy simulation presented in \autoref{sec:simulation}.
These properties can be studied either individually for every bubble or as a summarizing statistic for all bubbles in a specific snapshot.
We additionally compute characteristics that are known from observational studies, such as \citet{1997MNRAS.289..570O,pokhrel_catalog_2020}, and compare them to discuss similarities and differences between simulations and observations, taking into account the limitations of observations.

\subsection{Single bubble study}

We can trace the border particles 
of each hole (see \autoref{sec:bubbleorigine}) and
investigate their properties.
Here, we focus on three important physical properties of the SN-bubbles detected in a window of 11 snapshots (each 10 Myr apart) centred on snapshot 180: the expansion velocity $v^{\mathrm{exp}}$, the compactness $C$, and the gas temperature $T$.
 
To compute the expansion velocity, we first compute the mean velocity of all particles on the border of a hole, denoted as $\vec{\bar{v}} = \langle \vec{v}_i\rangle$, and subtract it from the velocity $\vec{v}_i$ of every particle:
\begin{equation}
    \hat{\vec{v}}_i = \vec{v}_i - \vec{\bar{v}}.
\end{equation}
Next, we compute the position of the centre of mass, $\vec{R}$, of all particles $\vec{b}_i$ on the border of the hole and calculate each border particle's projected velocity in the radial direction $v_i^{\mathrm{exp}}$ as
\begin{equation}
     v_i^{\mathrm{exp}} = \frac{(\vec{b}_i-\vec{R})\cdot \hat{\vec{v}}_i}{\lVert(\vec{b}_i-\vec{R})\rVert}. 
     \label{eq:Vexp0}
\end{equation}
The expansion velocity of the bubble, $\bar{v}^{\mathrm{exp}}$, is then obtained by averaging the radial velocities of all its $n$ border particles:
\begin{equation}
     \bar{v}^{\mathrm{exp}} = 
     \frac{1}{n} \sum_{i=1}^{n} v_i^{\mathrm{exp}} .
     \label{eq:Vexp}
\end{equation}
The compactness $C$ of a set of particles is defined as: \begin{equation}
 C = \frac{1}{\max_{i,j} \lVert \vec{b}_i - \vec{b}_j \rVert},
\end{equation}
where $\vec{b}_i$ and $\vec{b}_j$ are points on the border of the hole.  $C$  is
the inverse of the largest distance between pairs of border points, and thus provides  
a quantitative description of our intuitive notion of compactness. 
\def\mySNW{0.9}
\begin{figure}
  \centering
  \includegraphics[width=\mySNW\linewidth]{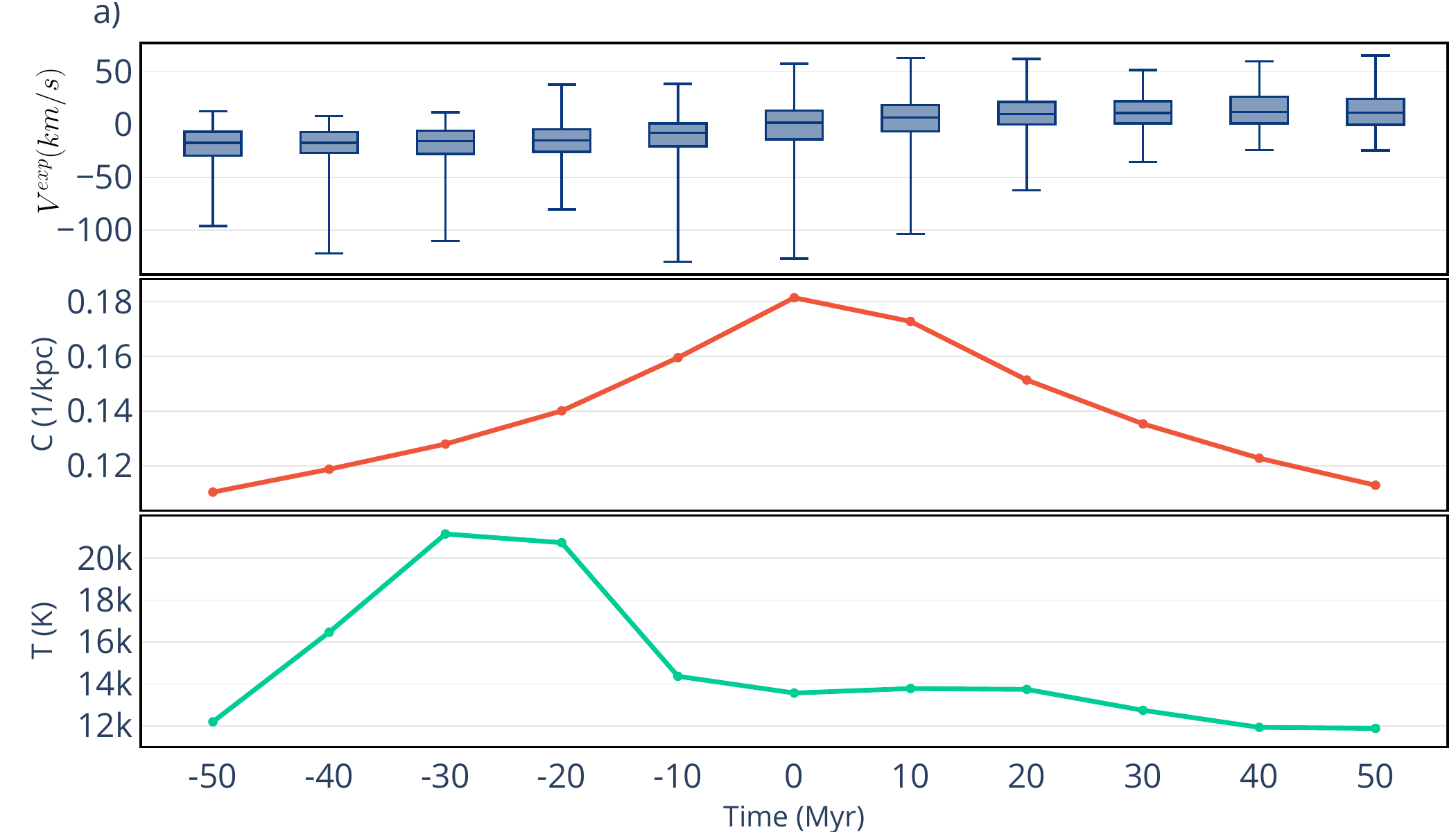}
  \includegraphics[width=\mySNW\linewidth]{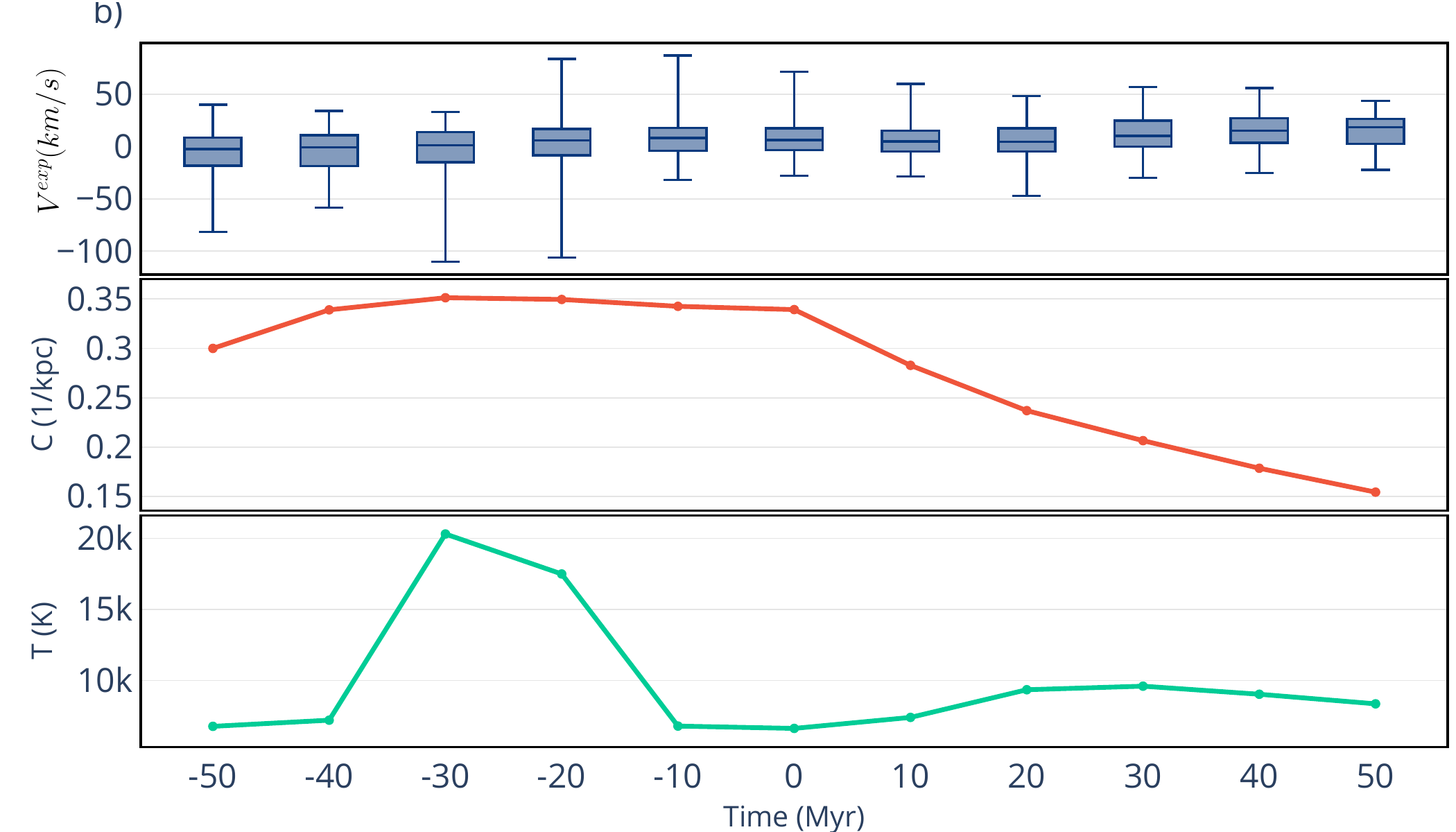}
  \includegraphics[width=\mySNW\linewidth]{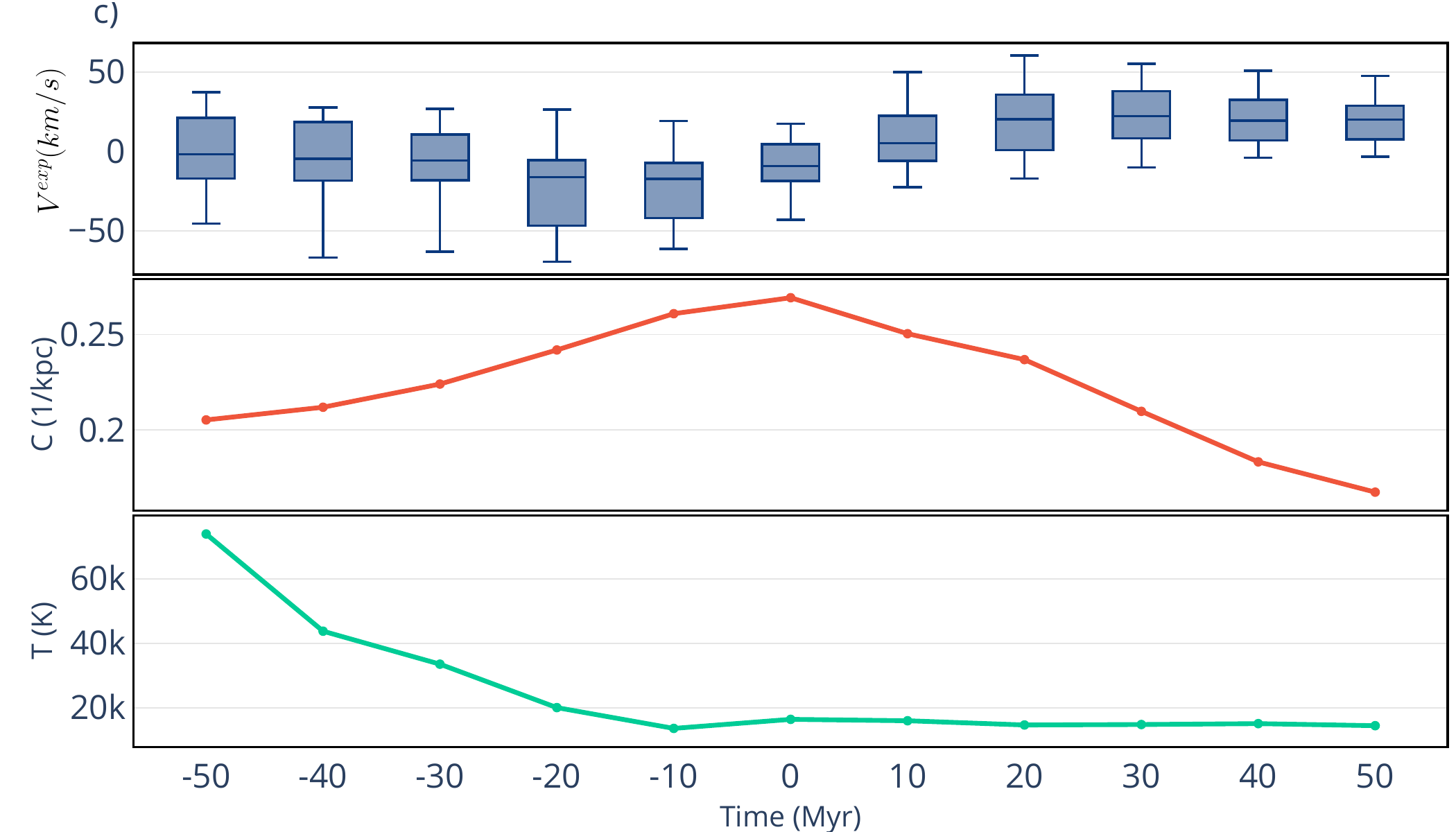}
  \includegraphics[width=\mySNW\linewidth]{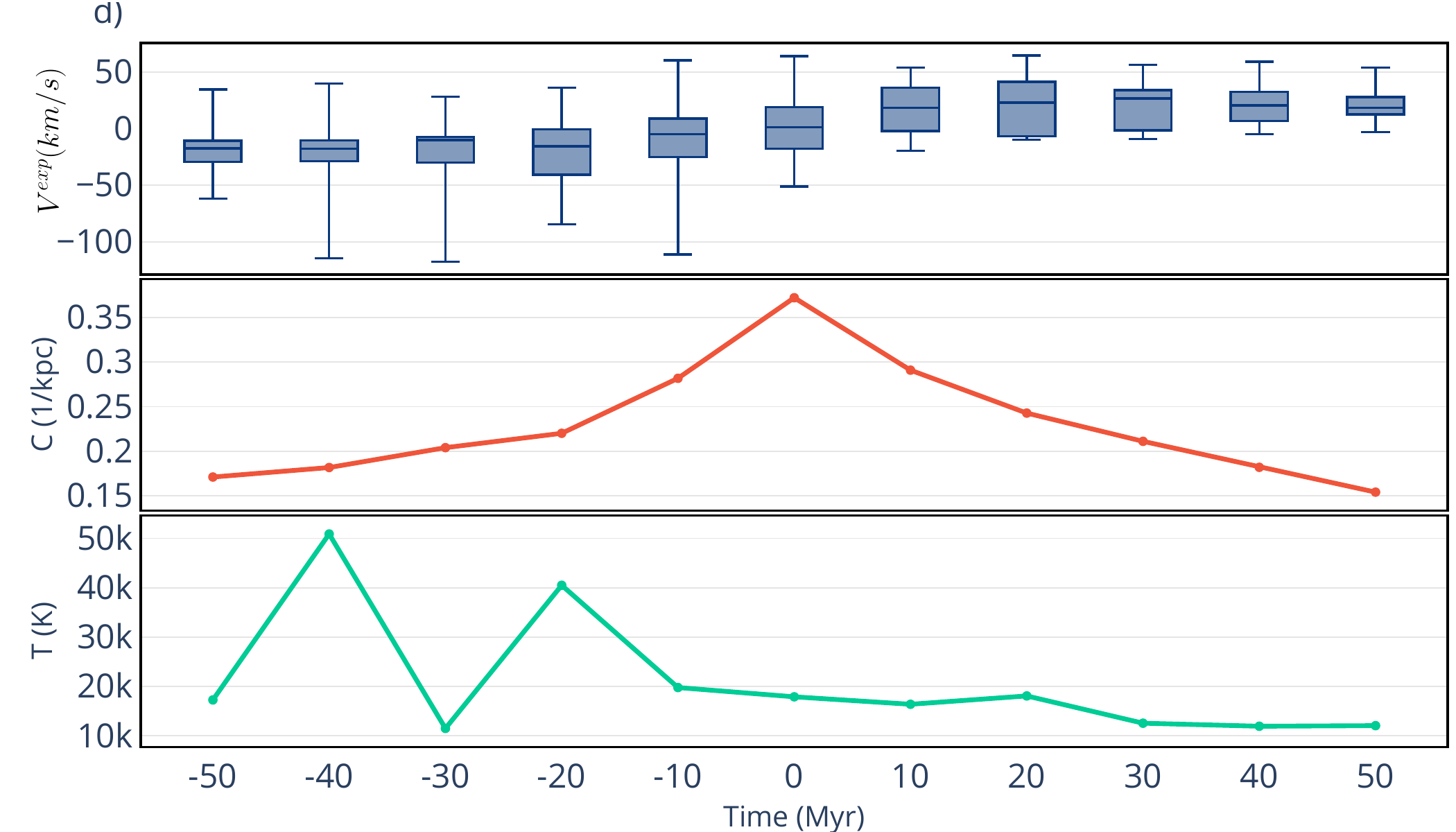}
\caption{\sven{The expansion velocity $v^{\mathrm{exp}}$, compactness $C$, and temperature $T$ of four SN-bubbles (a-d) detected in snapshot 180 (at Time~$=0$ in this figure), and their values in snapshots before and after detection. 
a) shows a typical SN-bubble, that contracts and exhibits a temperature surge $>15,000$~K, 30-40~Myr before its expansion.  
Not all SN-bubbles contract and expand, b) exemplifies a co-moving boundary with similar velocities and stable compactness in snapshots 176-180. Also, there can be more than one surge in temperature (see d).}} 
\label{fig:fourSN}
\end{figure}

\autoref{fig:fourSN} depicts the evolution of these three quantities in four SN-bubbles detected in snapshot 180. Contrary to the expectation, SN-bubbles are not necessarily expanding. 
Some cavities neither expand nor shrink, with particles on their boundary co-moving at similar velocities and maintaining stable compactness for several snapshots, as exemplified in \autoref{fig:fourSN}b). 
Furthermore, the mean temperature of particles on the border of an SN-bubble often rises a few snapshots earlier than the observed changes in mean expansion velocity and compactness. 
Additionally, \autoref{fig:fourSN}d) illustrates that more than one surge to a temperature greater than $15,000~K$ might occur.

Moreover, certain bubbles (such as in \autoref{fig:fourSN}d) might be short-lived, as indicated by a sharply peaked compactness at detection. Panels a) and c) of \autoref{fig:fourSN} also depict two SN-bubbles where, several snapshots after the temperature rise, particles are still contracting. Only after a few snapshots, i.e., after 10~Myr, do these bubbles start to expand, evidenced by an increasing expansion velocity and decreasing compactness.
\begin{figure*}
\centering 
\includegraphics[width=\linewidth]{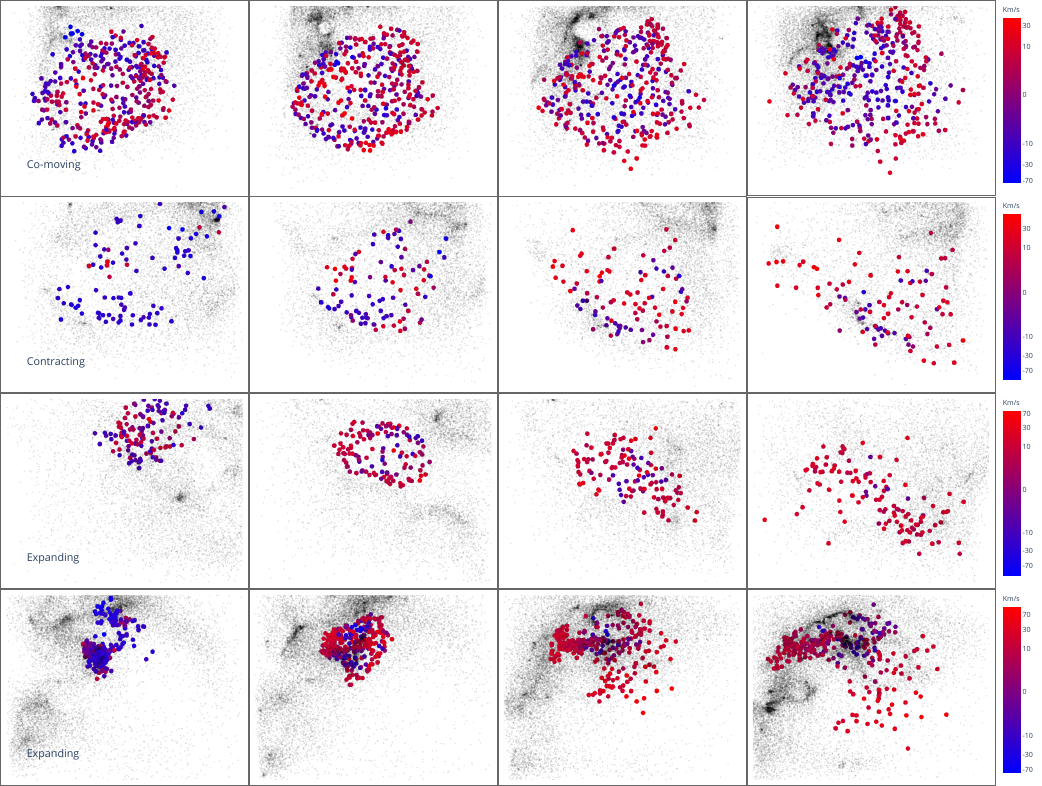}
\caption{The particles on the boundary of holes may co-move (top row), contract (second row), and expand in different directions (third and bottom rows). 
The colour of each  hull  particle  denotes the speed with which  
it is moving radially  away  
(red) or  towards 
(blue)  the bubble centre. Other particles are marked black, and each row shows a fixed size window that is not moving.
All 4 cavities were detected in snapshot 180 shown in the 2nd column.
The first column (snapshot 177) shows the same particles 30~Myrs before detection, while the 3rd and 4th (snapshots 183 and 186, respectively) illustrate the same particles (and their properties) in 30 and 60~Myrs after detection, respectively. The dimensions of the volume enclosing the supernova within each row exhibit variance. Sequentially, from the top to the bottom rows, these dimensions are as follows: $7\times8\times7$ kpc$^3$, $9\times6\times8$ kpc$^3$, $7\times4\times4$ kpc$^3$, and $6\times7\times6$ kpc$^3$.}
\label{fig:SNcandTypes} 
\end{figure*}
\autoref{fig:SNcandTypes} illustrates examples of border points of SN-bubbles, categorized as either co-moving (top row), contracting (second row), or expanding (third and bottom row). The colour of the data points indicates the speed and direction of the velocity of the border particles, with blue (red) data points corresponding to radially inward (outward) motion.

Each row displays four snapshots in the life of an SN-bubble detected in snapshot 180. 
The leftmost column shows a snapshot $30$~Myr prior to detection, the second column shows the bubble at the moment of detection, and the third and fourth columns show the bubble $30$ and $60$ Myr after detection, respectively.
The top row presents an example of a cavity that is co-moving, thus not expanding or shrinking.
The second row from 
top shows an SN-bubble with most of its boundary particles moving radially inwards, while the bottom two rows show cavities with a majority of expanding border particles.

Boundary points of cavities close to the centre of the jellyfish-like dwarf galaxy typically expand away from the centre, forming a cone-like structure away from the galaxy.
In that region of the jellyfish, many activities and explosions occur in close vicinity and a similar time frame, indicating that bubbles are not only affected by supernova explosions inside them but also by other feedback mechanisms. This observation contrasts with \citet{pokhrel_catalog_2020}, where it was assumed that the cavities in all studied dwarf galaxies are expanding.
Up to this point, we designate a bubble as expanding (contracting) if more than 50\% of particles on its border are moving away from (towards) its centre. Otherwise, the cavity is labelled as co-moving. 
As this threshold is somewhat arbitrary, we repeat the counts with increasingly strict thresholds for the fraction of expanding or contracting particles, resulting in an increased number of co-moving holes. 
\autoref{tab:expcon} presents the number of observed holes or SN-bubbles in snapshots 170 to 190 that are expanding, contracting, or co-moving, given a threshold on the fraction of particles exhibiting the corresponding behaviour.
 
\begin{table}
\centering
\caption{Number of SN-bubbles and nonSN-bubbles in snapshots 170 to 190 with hull particle behaviour given different thresholds (see text for details).}
\label{tab:expcon}
\begin{tabularx}{0.85\columnwidth}{@{\extracolsep{\fill}} cl|rrr}
\toprule
& & \makecell[c]{Expanding} & \makecell[c]{Contracting} & \makecell[c]{Co-moving}\\
\hline 
 \multirow{3}{*}{\rotatebox[origin=c]{90}{\makecell{nonSN-\\bubbles}}}
 & > 50\% & 477 & 293 & 38 \\
 & > 60\% & 294 & 130 & 384\\
 & > 70\% & 147 & 60 & 601 \\
 \hline 
 \multirow{3}{*}{\rotatebox[origin=c]{90}{\makecell{SN-\\
 bubble}}}
 & > 50\% & 245 & 45 & 6 \\
 & > 60\% & 192 & 14 & 90 \\
 & > 70\% & 117 & 5 & 174 \\
\bottomrule
\end{tabularx}
\end{table}

\subsection{Statistical information about the detected cavities}
\label{sec:statistics}

The properties of a discovered hole vary in different snapshots of the particle simulation. 
Note, that our technical contribution \cite{taghribi_asap_2021} demonstrates the detection and modeling of arbitrary general shapes of bubbles, that allow extraction of realistic volumetric information. 
The approach builds a probabilistic model of the recovered cavity using the methodology of (\cite{canducci_probabilistic_2022}). This in turn leads to the construction of appropriate likelihood iso-surfaces.
However, to the best of our knowledge, observational studies typically report measurements based on simple geometrical shapes (such as circles) in their descriptions of cavities.
In this section, we therefore present a statistical analysis of the diameter, estimated age, and the released energy for every non-SN cavity and SN-bubble. 
These properties are comparable to the characteristics investigated in \citet{pokhrel_catalog_2020} for 41 observed dwarf galaxies, and the results show similarities in the analysed properties in both simulation and observation.

One approach to computing properties such as age and released energy is to directly obtain this information from the simulation. To identify whether a type II supernova (or \snii{}) caused a cavity, we verify the existence of one or more stellar particles with active \snii{} inside it. To account for the time delay between the explosion of a \snii{} and the emergence of a cavity, we proceed as follows for each detected cavity. First, we define a search sphere centred on the cavity, with a radius equal to that of the cavity at the time of detection. Second, we trace the particles on the boundary of the cavity back through time to recover their locations in the snapshot where their average temperature first exceeds 15,000~K, and we compute the average of their positions. Third, we translate the search sphere to that average position, keeping its radius fixed. Finally, we search for stellar particles within the search sphere that are actively injecting energy into the surrounding gas.

\begin{figure}
\centering
\includegraphics[width=1\linewidth]{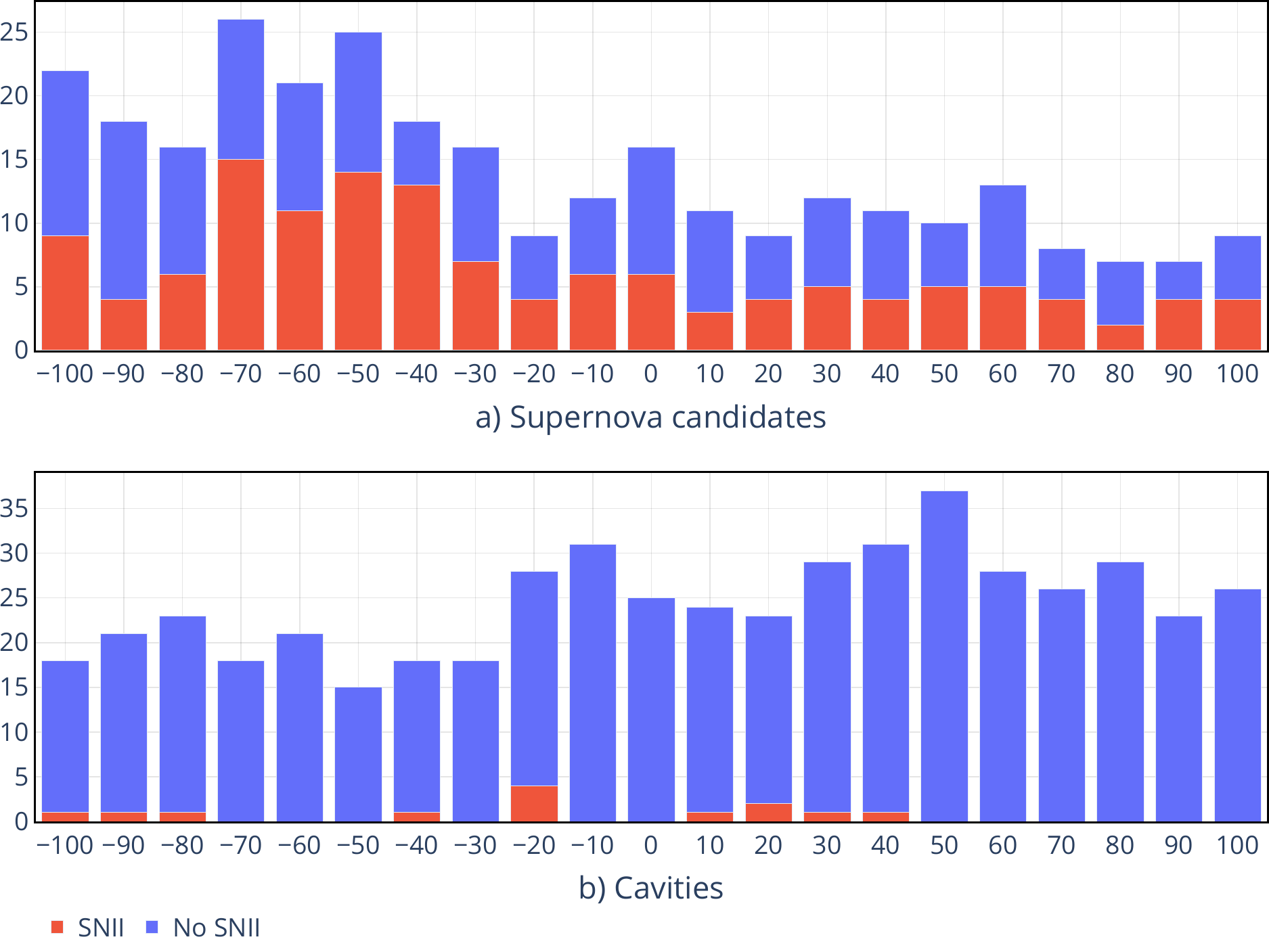}
\caption{A comparison of the fraction of candidate SN-bubbles (top row) and nonSN-bubbles (bottom row), identified using the criteria listed in \autoref{sec:bubbleorigine}, that contain \snii\ particles, identified using the procedure from \autoref{sec:statistics}. Clearly, the simple SN-bubble detection criteria from \autoref{sec:bubbleorigine} perform well: only a small fraction of the nonSN-bubbles actually contain stellar particles that actively give feedback to the surrounding ISM while over half of the candidate SN-bubbles contain detected \snii\ particles.
}
\label{fig:holeSNII}
\end{figure}

\autoref{fig:holeSNII} illustrates how many of the candidate SN-bubbles and nonSN-bubbles identified in a snapshot contain more than one \snii{} particle, detected using the procedure described above. In all snapshots, significantly more \snii{} are observed in candidate SN-bubbles compared to nonSN-bubbles, which do not satisfy the conditions introduced in \autoref{sec:bubbleorigine}. This shows that these conditions provide a reliable criterion for rapidly and straightforwardly detecting gas bubbles inflated by stellar feedback. It should be noted that due to the haphazard motion of the gas and star particles between the two snapshots involved in this procedure, there is no unique or optimal way of defining the search sphere. As a result, the search sphere may miss some \snii{} particles that actually helped inflate the bubble, or, conversely, erroneously include \snii{} that are unrelated to the bubble.

\subsubsection{Diameter of nonSN-bubbles and SN-bubbles}  

We subsampled the data 100 times using \textsc{asap} (see \autoref{sec:holelocation}) and studied the persistent homology of its features built on alpha filtration to find robust cavities. 
Subsequently, the mean 
radius (or diameter) is calculated over 100 repeated detections 
with the simplified assumption of spherical cavities.

In the appendix, \autoref{fig:diameter} shows histograms of the diameter of all holes identified (including SN-bubbles) in every snapshot (in blue) and of the subset of only SN-bubbles (in red). In the snapshots, the diameters of all cavities range between 0.66~kpc and 9.06~kpc. The diameter of the SN-bubbles, however, is smaller, ranging only from 0.66~kpc to 5.54~kpc. Since the subsampling with \textsc{asap} used a radius of 0.15~kpc, every hole with a diameter smaller than this is statistically insignificant. The detection limit depends on the data resolution, as discussed in more detail in appendix \ref{sec:data limits}.

It is noteworthy that while the reported diameter range for SN-bubbles exceeds the expected range from observation-based studies (0.2 to 2~kpc as per \citet{pokhrel_catalog_2020}), 90\% of detected SN-bubbles exhibit a diameter below 2.98~kpc. The median detected diameter is 1.17~kpc, aligning more closely with other studies.

\subsubsection{Expansion velocity}
\label{sec:Vexp}
For every particle on the border of an identified hole, the  hull-particle expansion velocity is calculated according to equation~\eqref{eq:Vexp} in the same snapshot it
was detected.

Appendix \autoref{fig:Vexp} shows histograms of the mean expansion velocity for all holes in every snapshot, which varies between $-26.48$~km/s and $47.35$~km/s. For some bubbles, the mean expansion velocity is negative, indicating that the cavity is contracting, which does not conform to the expansion assumption in \citet{pokhrel_catalog_2020}. Many of these holes appear in a turbulent area of the galaxy, and other supernova explosions or gas disruption may affect the boundary particles and influence the direction of their movement. 
Consequently, the assumption of a positive mean expansion velocity does not hold for all SN-bubbles and \mc{cavities} 
discovered in the dwarf galaxy simulation. 
Furthermore, we note that the mean expansion velocity of SN-bubbles is usually higher than for cavities not classified as SN-bubbles. 

\subsubsection{The kinetic age}

\cite{pokhrel_catalog_2020} suggest computing the kinetic age of a cavity by assuming a constant velocity similar for all border particles, a constant rate of expansion during its lifetime, and a spherical shape. Hence, the kinematic age is given by the following equation:
\begin{equation}
    t_{\mathrm{kin}} = 0.978\frac{d/2}{| \bar{v}^\mathrm{exp}|} \enspace ,
    \label{eq:tkin}
\end{equation}
where $d$ is the diameter and $|\bar{v}^\mathrm{exp}|$ is the expansion velocity of the hole. However, as we discussed in \autoref{sec:Vexp}, we know from the simulation that the expansion velocity is different for every particle on the border of a cavity, and the mean expansion velocity might even be negative. Therefore, equation~\eqref{eq:tkin} is not expected to accurately estimate the age of our cavities. Therefore, we propose another way to estimate the bubble's age in the simulation. As explained in \autoref{sec:bubbleorigine}, one of the signs to distinguish a supernova explosion from a bubble caused by other physical phenomena is the rise of the mean temperature of the particles on the border of the hole beyond 15,000~K. The increase in temperature precedes other signs of a supernova explosion and, hence, can reach a better estimate of its kinetic age.

If we assume that the snapshot with the highest mean temperature is the one when the explosion starts, we can count the number of snapshots between the starting point and the detection of the cavity, knowing the difference between two consecutive snapshots is 10 million years in this simulation. Of course, it is only possible to compute the age as a multiple of 10 million years in this scenario.

We are well aware that not all cavities are caused by supernova explosions, and hence both the estimation of the kinetic age using equation~\eqref{eq:tkin} and our strategy based on the highest mean temperature are limited in their accuracy. 
Nevertheless, for the SN-bubbles, these two strategies still provide relevant information. 
The appendix \autoref{fig:age_abs} shows the histogram of the kinetic age of holes (blue) and the subset of SN-bubbles (red) based on the mean absolute expansion velocity of all border particles and equation~\eqref{eq:tkin}. 
This figure is comparable to the ages reported in \cite{pokhrel_catalog_2020}. 
The SN-bubbles' age is mostly less than 100 million years, while  
the age of a cavity 
not identified as an SN-bubble could be more than 400 million years.

In contrast, the histograms of the age estimated using the number of snapshots after the highest mean temperature rise as described before result in ages of an SN-bubble mostly varying between 0 and 50 million years (see \autoref{fig:age_close} in the appendix). Note that there are also many holes and a few SN-bubbles with a negative age, for which the snapshot with the maximum mean temperature happens after detection. As shown in \autoref{fig:fourSN}(d), there might also be multiple peaks in the temperature profile, and we compute the age based on the one with the higher temperature value. This may also lead to negative numbers for the kinetic age, which shows the limitations of estimates based solely on the topological features built by the particles and that caution also in the interpretation of observations may be advised.

\subsubsection{Energy}
Two different ways of computing energy were suggested in the literature. 
The first technique proposed by \cite{chevalier_evolution_1974} 
assumes a single explosion shaped the supernova and the released energy is estimated by: 
\begin{equation}
    E_\mathrm{Ch} = 5.3\cdot 10^{43} \rho^{1.12} \left(\frac{d}{2}\right)^{3.12} |\bar{v}^\mathrm{exp}|^{1.4} \enspace .
    \label{eq:energy_74}
\end{equation}
Here, the metrics for the \mc{energy $E_\mathrm{Ch}$, density $\rho$ of gas, diameter $d$, and} $|\bar{v}^\mathrm{exp}|$ are erg, cm$^{-3}$, pc, and km/s, respectively. 
The second way suggested by \cite{mccray_supershells_1987} assumes multiple supernovae explosions created the cavity:
\begin{equation}
E_\mathrm{Mc} = \rho \left(\frac{d}{194}\right)^2 
\left(\frac{|\bar{v}^\mathrm{exp}|}{5.7}\right)^3\cdot10^{51} \enspace .
\label{eq:energy_87}
\end{equation}
However, 
equations \eqref{eq:energy_74} and \eqref{eq:energy_87} 
assume that the velocity of expansion is always positive, which is not always the case. Hence, we again estimate the $|\bar{v}^\mathrm{exp}|$ by the mean absolute value of expansion velocity over all particles detected on the border of a hole. 
When estimating the energy released computed with equation~\eqref{eq:energy_74} and \eqref{eq:energy_87}, as expected,
the required amount of energy to form a cavity caused 
by a supernova explosion is higher than for other types of holes.
The respective figures \ref{fig:energy_74} and \ref{fig:energy_87} can be found in the appendix.

\section{Conclusions}
\label{sec:conclusion}
In this study, we demonstrate an automated pipeline for discovering cavities in a simulation of a jellyfish-like dwarf galaxy. 
The pipeline consists of a sub-sampling step that preserves topological features of the data with \textsc{asap}, then discovers the cavities using persistent homology and identifies their boundary points robustly, based on repetition and voting. 
By following these points or gas particles in consecutive snapshots of the simulation, one can analyse the properties of the cavities 
including the temperature, released energy, and age, and compare them with relevant studies on observational data, such as \cite{pokhrel_catalog_2020}.


We identified 808 holes in snapshots 170 to 190 (including both starting and ending snapshots). 
Only 296 of these cavities are interpreted as SN-bubbles (i.e. the mean temperature of particles on their border surpasses 15000~K and the orientation of more particles alters toward outside the centre). 
Furthermore, over 36\% of the holes are shrinking, not expanding. which contrasts
the assumption that all cavities are supernova-blown bubbles and, therefore, are expanding \citep{pokhrel_catalog_2020}. 
Despite the fact that it is observationally impossible to distinguish a shrinking from an expanding bubble, the assumption that all cavities expand has been central to observational studies of the interstellar medium of dwarf galaxies. 
Moreover, we show that, in different stages, a bubble may shrink and then start to expand, or some particles move towards the centre and others away from it. 
We also noticed 
that the temperature increase and the change from shrinking to expanding does not happen simultaneously.


In the second part of the study (\autoref{sec:bubbleorigine}) we explore the possibility of recovering the connection between known \snii\ particles in the simulation and detected cavities, purely based on topological features. 
As shown in \autoref{sec:statistics}, there is a delay between the mass loss process of a star particle and an increase in the temperature of surrounding gas particles. 
Hence, we only select the \snii\ particles located inside a cavity. 
Besides, the mass loss of \snii\ and increase in the temperature of gas particles on the boundary of the hole should occur concurrently.
Typically a cavity is assumed to be spherical (when followed back in time), to compute certain properties, such as diameter. 
Due to strong movement of the particles in our turbulent dwarf galaxy environment this is not \mc{completely} reliable, but still provides useful insights into the simulation.
The holes we classified as SN-bubbles are indeed more probable to contain \snii.
Furthermore, many holes do not contain any \snii\ while the others may include up to 100 \snii\ using the described matching process. 

Finally, we explored different properties of the robustly detected cavities 
in 21 snapshots.
The SN-bubbles typically have a smaller diameter (0-4~kpc), larger expansion velocity, and shorter kinetic age than other types of holes. 
We discuss an alternative approach to estimating the age of a hole 
by computing the time difference between the time with the highest average temperature of hull particles 
and the detection snapshot. 
This study shows that the age for many SN-bubbles is between 0 to 50 million years. 
Furthermore, the SN-bubbles in all snapshots released a higher amount of energy than other types of holes.
The radius of detected bubbles 
in this simulation was larger than 0.3 kpc, which contrasts expectations 
from observation data, that suggests smaller radii of 0.1 and 0.2 kpc for cavities caused 
by a supernova. 
Independent of the methodology, there is a detection limit on the size of holes due to 
the particle density in the data. 
If it is undercut very small cavities cannot robustly be distinguished from spurious holes.
For our proposed pipeline we demonstrate in 
appendix \ref{sec:data limits} that bubbles with a radius of 0.1~kpc require
a simulation with at least 4261 particles per cubic kpc. 

\section*{Acknowledgements}
We acknowledge financial support from the European Union's Horizon 2020 research and innovation programme under the Marie Sk\l odowska-Curie
grant agreement N.~721463 to the SUNDIAL ITN network.

\section*{Data Availability}
The data underlying this article is shared along the implementation at: 
\url{https://github.com/abst0603/ASAP}.



\bibliographystyle{mnras}
\bibliography{supernovae} 




\appendix

\section{Data requirements for the detection of 0.1 kpc bubbles} 
\label{sec:data limits}

\begin{figure*}
\centering
\includegraphics[width=0.95\linewidth]{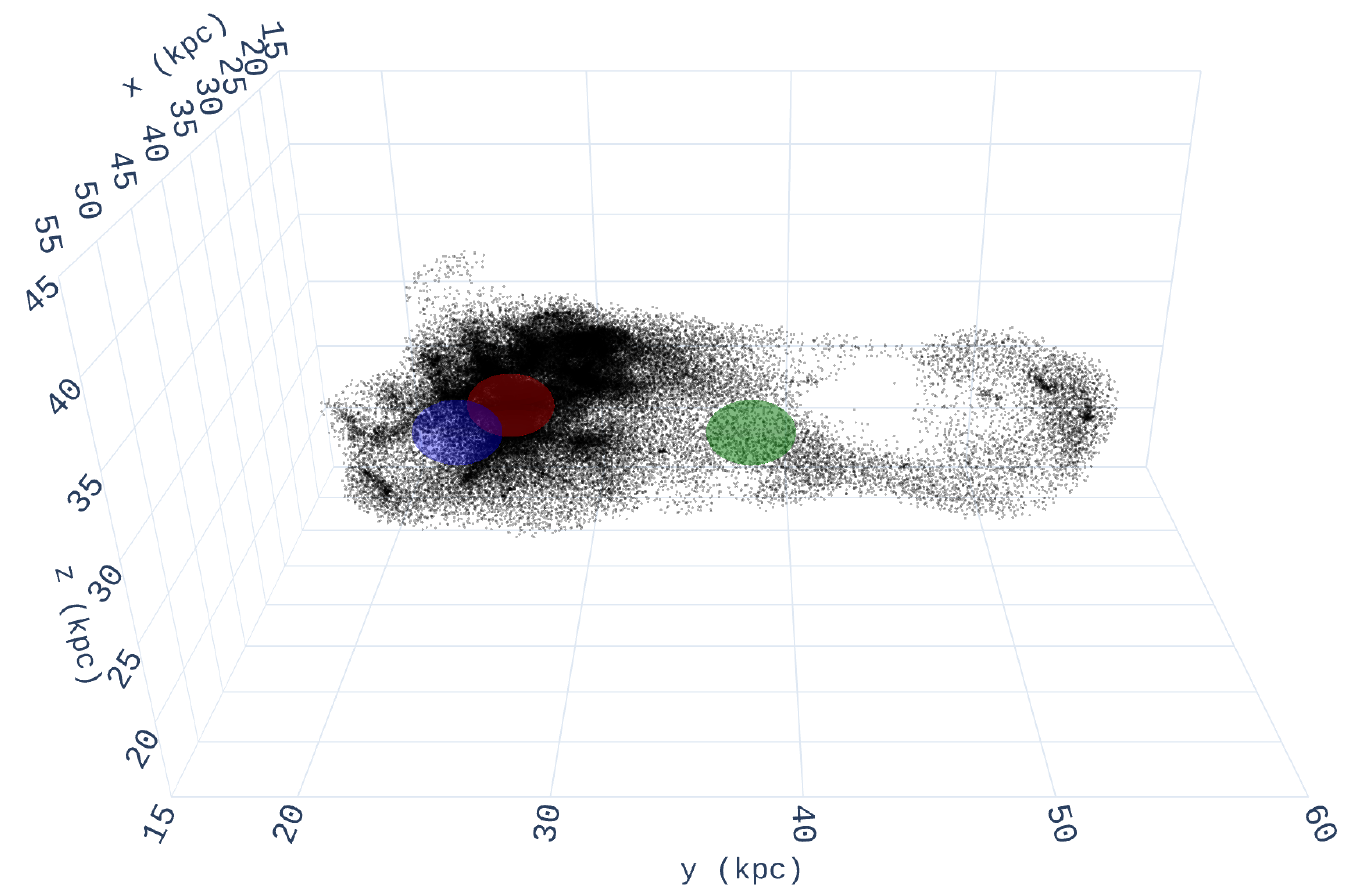}
\caption{Three spheres in red, blue, and green demonstrate three regions inside the jellyfish-like dwarf galaxy simulation with high, medium, and low density. 
The radius of all spheres is 
2~kpc and the axis unit is kpc.}
\label{fig:densitysamples}
\end{figure*}

Comparing the distribution of the diameter of detected holes in this study with the observation data~\citep{pokhrel_catalog_2020} reveals that the diameters of detected holes in this simulation (
\autoref{fig:diameter}) are generally larger than the expected diameter size of a supernova (between 0.2 to 2 kpc). 
In this section, we elucidate that the larger diameters are not a general problem of our detection strategy, but rather caused by limitations in the resolution (local density of particles) in the simulation this study is based on.
To validate this claim, we create three mock datasets 
resembling the density of a different region of the simulation, which are marked by three spheres in 
\autoref{fig:densitysamples}. 
The red sphere close to the centre of the jellyfish head (highest density), the blue one close to the border of its head (medium density), and the green one toward the jellyfish tails (lowest density).
The corresponding particle counts can be found 
in \autoref{table:density}.
The density in this section refers to the number of particles in $1$~kpc$^3$, not the gas density $\rho$, which is a parameter in simulation. 
\begin{table}
\centering
\caption{The density of samples taken from three regions inside snapshot 180 of the simulation, as illustrated in 
\autoref{fig:densitysamples}.}
\label{table:density}
\begin{tabularx}{0.85\columnwidth}{@{\extracolsep{\fill}} cc|c}
\toprule
& & \makecell[c]{Density (number of particles per cubic kpc)} \\
\hline
 \multirow{3}{*}{\rotatebox[origin=c]{90}{Samples}}
 & Red & 284.09 \\
 & Blue & 41.09 \\
 & Green & 5.22 \\
\bottomrule
\end{tabularx}
\end{table}

\begin{figure}
\centering
\includegraphics[height=0.9\linewidth]{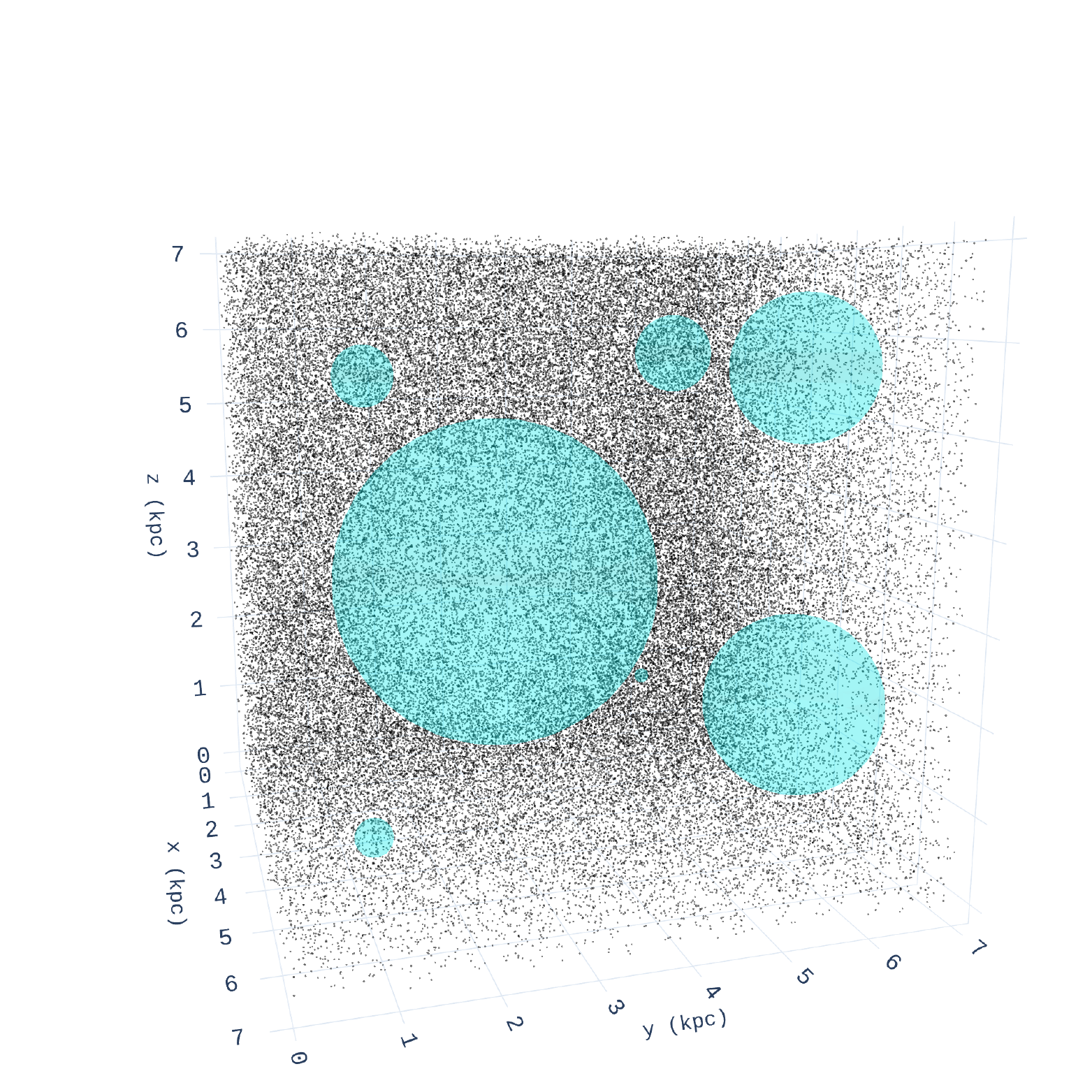}
\caption{
Synthetic dataset with a uniform point cloud resembling the highest density found in our simulation data 
(corresponding to the red sphere in 
\autoref{fig:densitysamples}) and seven cavities of different radii extracted (cyan holes).}
\label{fig:mockdataset}
\end{figure}

We generate three mock datasets with uniformly distributed points according to the sampled densities, containing seven cavities 
with various radii in kpc:  [0.1,0.2,0.3,0.5,0.8,1,2].
The mock dataset with the highest density is visualized in 
\autoref{fig:mockdataset}.
The same pipeline explained in 
\autoref{sec:holedetection} was applied to these mock datasets to recover the holes. 
Within the lowest density cube, we robustly 
detect the two largest holes (1~kpc and 2~kpc) and identify the particles on their boundaries. 
For the other two denser datasets, the five largest holes were detected. 
However, the two cavities with small radii (0.1~kpc and 0.2~kpc) were not recovered clearly in 
their assigned space.
It is worth mentioning, that several holes with a radius close to 0.2~kpc were recovered in both denser datasets, but those 
are spurious random patterns caused by the intra-particle distances. 
Hence, in our simulation with the given density limitations we cannot discriminate between holes created by meaningful physical features and topological noise when they are small.
Subsequently, to identify supernovae with a radius close to 0.1~kpc, a simulation with a higher density of particles is required.

\begin{figure}
\centering
\includegraphics[width=1\linewidth]{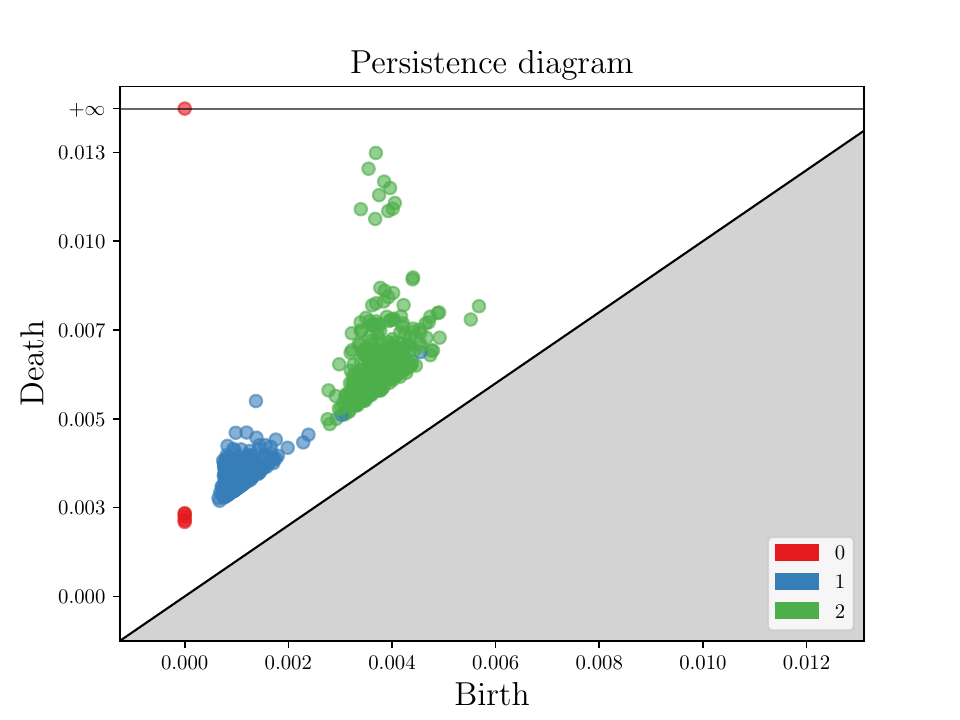}
\caption{Persistent homology plot for a dataset with a density of 4261 particles in a unit kpc cube, that is sufficient to detect
and distinguish the ten 0.1~kpc radius ground truth cavities from noisy features. 
Their death time approximates the squared radius of the created holes.
}
\label{fig:PHproperdensity}
\end{figure}

To understand the required density necessary to recover holes with a radius of 0.1~kpc, we generated another mock dataset containing ten 0.1~kpc radius cavities. 
Starting from the highest density, as approximated from the dwarf jellyfish galaxy simulation, we 
gradually increased (multiplies of starting density) the density of uniform points in the cube surrounding the 10 cavities. 
Every step 
we sketch the persistent homology plot for the dataset to scrutinize if all ten holes can be isolated from noise patterns. 
We observe that in a dataset with 15 times more particles (4261 particles per cubic kpc) than the highest density sampled from dwarf jellyfish galaxy simulation, we can robustly differentiate between created ground truth cavities and random 
topological features exhibiting lower persistence.
This is indicated by the ten green points representing the birth and death time of each 
of the ten holes created inside the dataset, that are well separated from the other more noisy features shown 
in \autoref{fig:PHproperdensity}. 
The red and blue points show the representations of lower dimensional features with betti number 0 (connected components) and 1 (cycles) inside the dataset, that are not further discussed here.
Due to generating the data using uniform sampling and then removing particles inside the bubbles, the radius of each bubble can be slightly larger. 
As presented in 
\autoref{fig:PHproperdensity} the death time of the ten generated holes is between 0.01 to 0.013, and they are evidently separated from the rest of the topological features. 
Moreover, since alpha filtration is applied 
to compute the persistent homology, the death time is close to the expected radius squared for created features inside the dataset.

\section{Histograms of bubble properties}

\begin{figure*}
\centering
\includegraphics[width=1\linewidth]{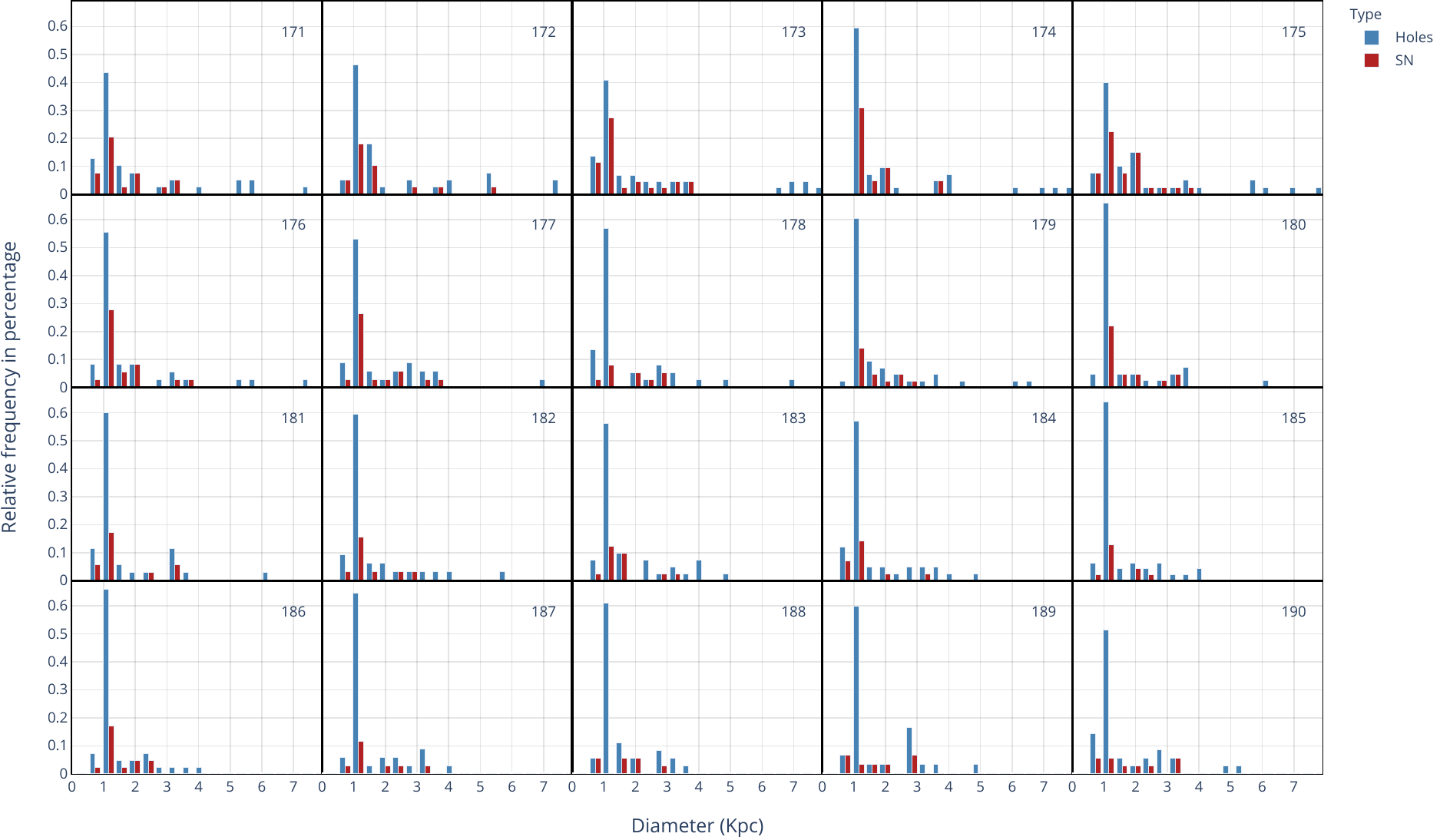}
\caption{Histograms of estimated 
diameter of all holes (blue) and the subset of} SN-bubbles (red) in every snapshot (171-190).
\label{fig:diameter}
\end{figure*}

\begin{figure*}
\centering
\includegraphics[width=1\linewidth]{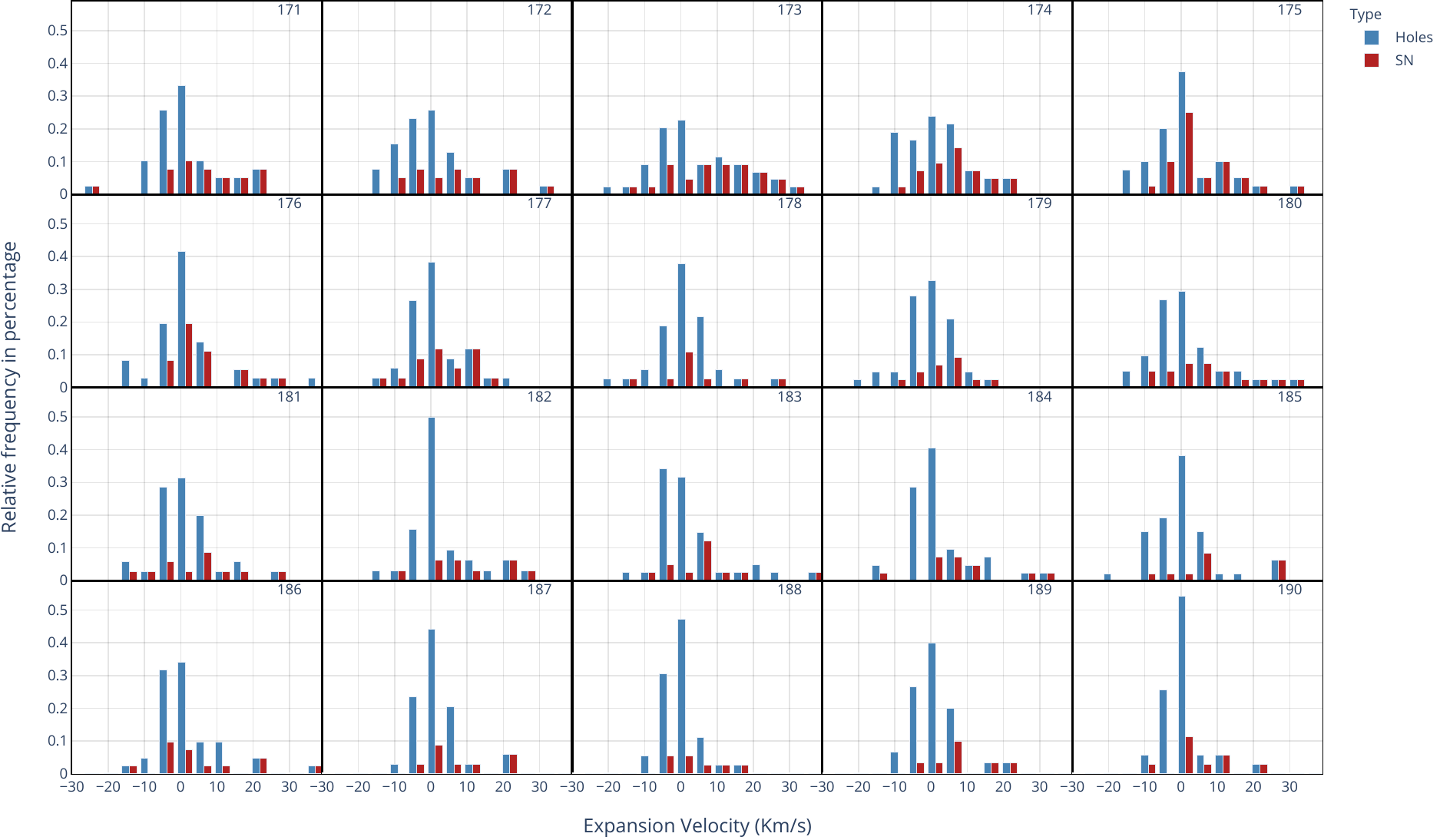}
\caption{Histograms of mean expansion velocity of boundary particles of holes (blue) the subset of SN-bubbles (red) in every snapshot (171-190).}
\label{fig:Vexp}
\end{figure*}

\begin{figure*}
\centering
\includegraphics[width=1\linewidth]{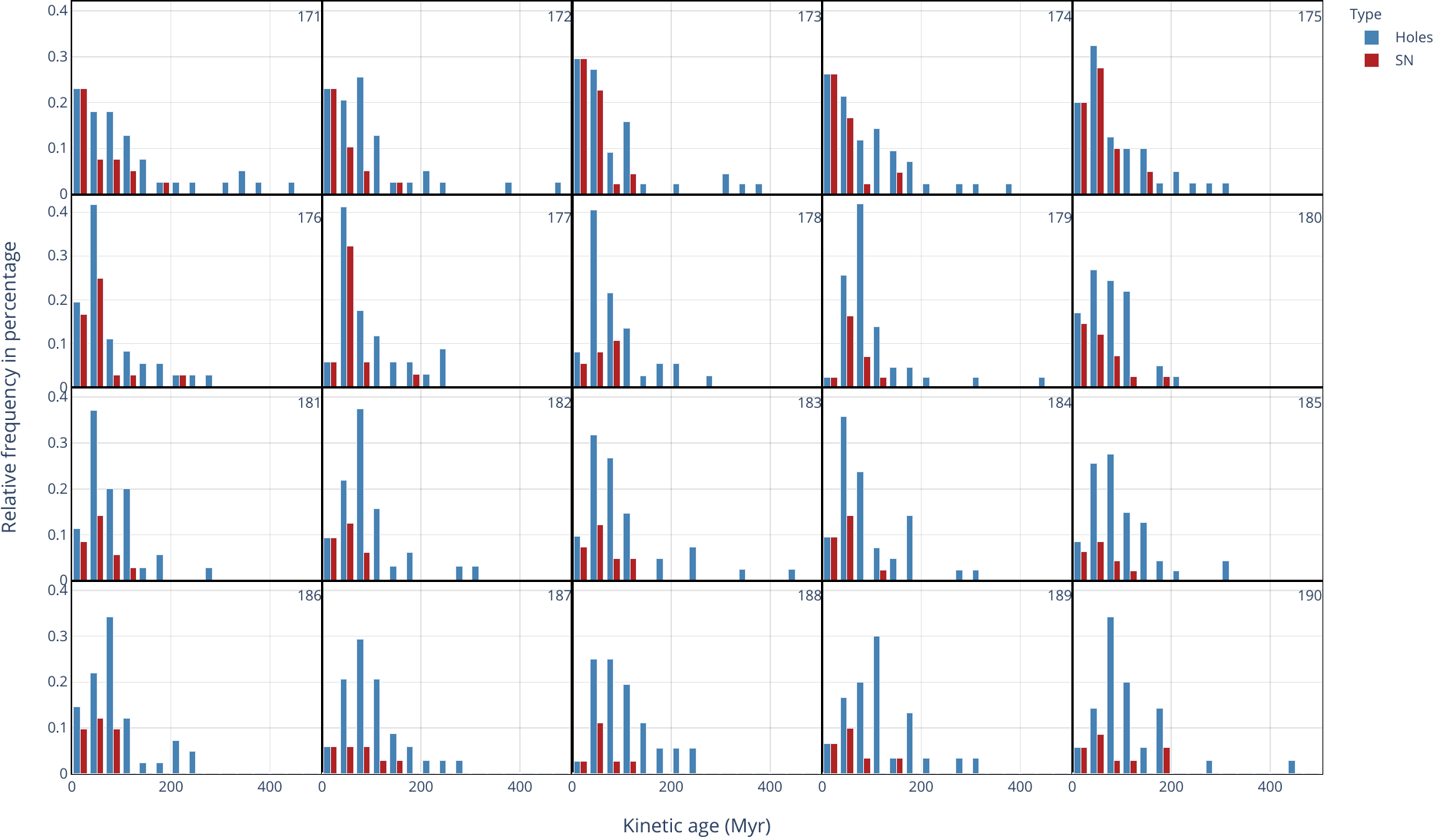}
\caption{Histograms of kinetic ages of holes (blue) and the subset of SN-bubbles (red) in our simulation computed with 
equation~\eqref{eq:tkin} in all snapshots.}
\label{fig:age_abs}
\end{figure*}

\begin{figure*}
\centering
\includegraphics[width=1\linewidth]{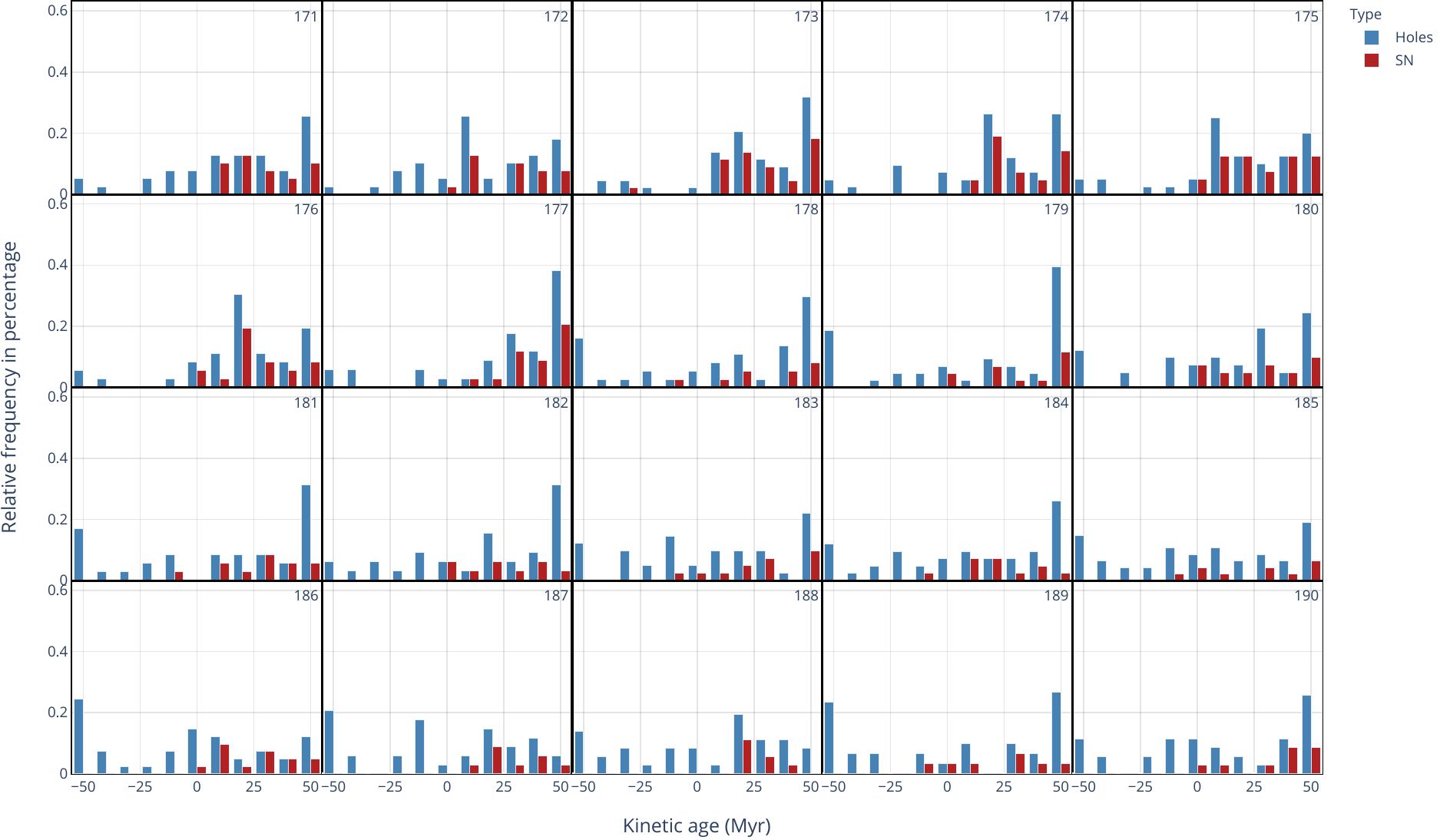}
\caption{Histograms of the kinetic age of holes (blue) and the subset of SN-bubbles(red) based on the maximum mean temperature rise (frame 
171 to 190).}
\label{fig:age_close}
\end{figure*}

\begin{figure*}
\centering
\includegraphics[width=1\linewidth]{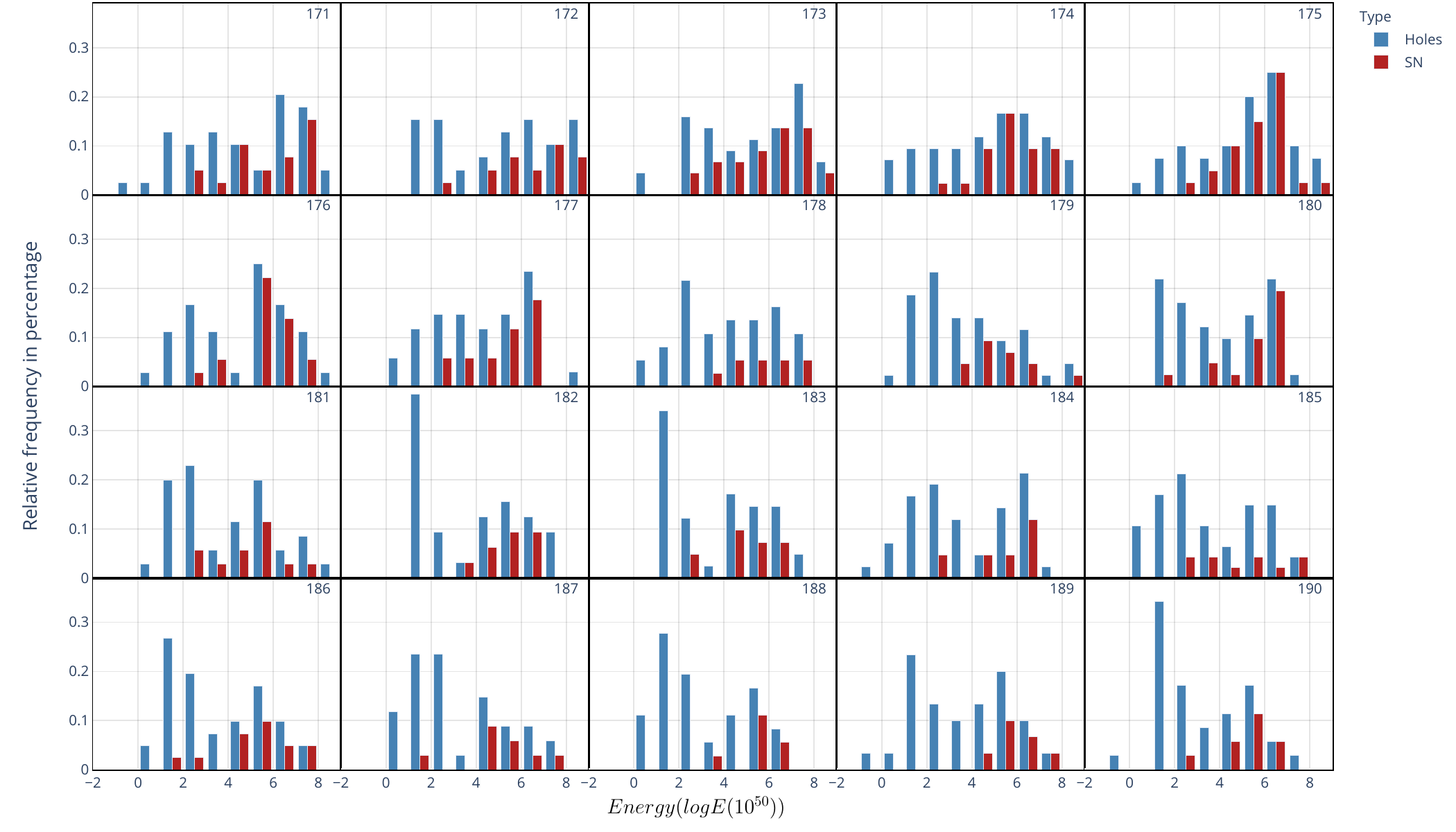}
\caption{Histograms of holes (blue) and the subset of SN-bubbles' (red) released energy computed using equation~\eqref{eq:energy_74} in all snapshots.}
\label{fig:energy_74}
\end{figure*}

\begin{figure*}
\centering
\includegraphics[width=1\linewidth]{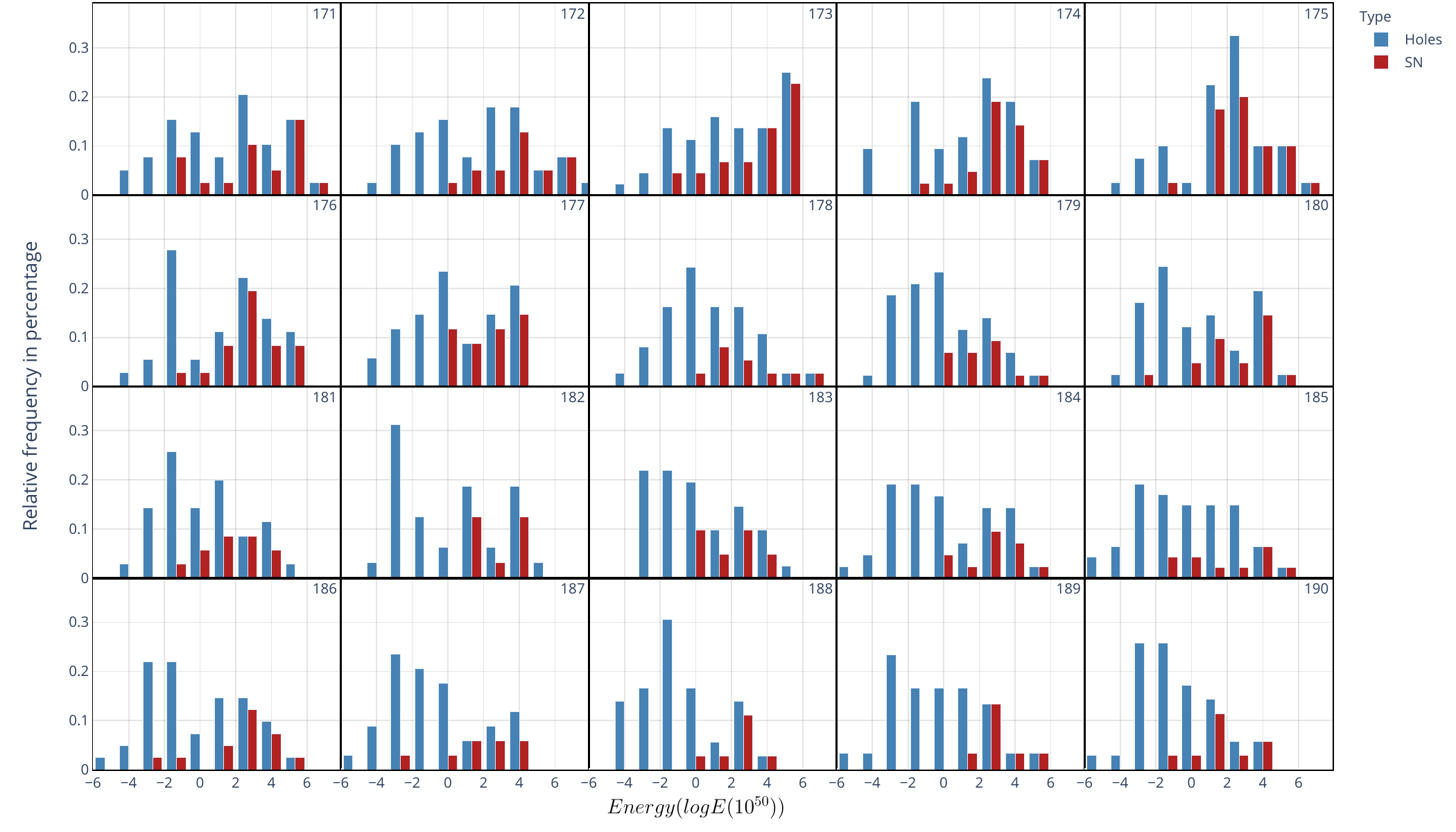}
\caption{Histograms of holes (blue) and the subset of SN-bubbles' (red) released energy computed using equation~\eqref{eq:energy_87} in all snapshots.}
\label{fig:energy_87}
\end{figure*}

\autoref{fig:diameter} shows histograms of the 
diameter of all holes identified (including SN-bubbles) in every snapshot (in blue) and the subset of SN-bubbles (in red). 
In the snapshots, the diameters of all cavities range 
between 0.66~kpc and 9.06~kpc. The diameter of the SN-bubbles, however, is smaller, ranging only from 
0.66~kpc to 5.54~kpc. 
\autoref{fig:Vexp} shows histograms of the mean expansion velocity for all holes in every snapshot, which varies between $-26.48$~km/s to $47.35$~km/s. 
For some bubbles and SN-bubbles the mean expansion velocity is negative, indicating that the cavity is contracting, which does not conform to the 
expansion assumption in \citet{pokhrel_catalog_2020}. 
\autoref{fig:age_abs} shows the histogram of the kinetic age of holes (blue) and the subset of SN-bubbles (red) based on the mean absolute expansion velocity of all border particles and
equation~\eqref{eq:tkin}. 
This figure is comparable to the ages reported in \cite{pokhrel_catalog_2020}. 
The SN-bubbles age are mostly less than 100 million years, however, the age of a cavity that is not identified as a SN-bubble could be more than 400 million years.
\autoref{fig:age_close} shows the histograms of the age estimated using the number of snapshots after the highest mean temperature rise as described before, resulting in
ages of a SN-bubble mostly varying 
between 0 and 50 million years. 
Note that, there are also many holes and a few SN-bubbles with a negative age, for which 
the snapshot with the maximum mean temperature happens after detection.
\autoref{fig:energy_74} and \ref{fig:energy_87} show the estimated energy released computed with equation~\eqref{eq:energy_74} and \eqref{eq:energy_87}, respectively.
In both figures, the required amount of energy to form a cavity caused 
by a supernova explosion is higher than for other types of holes.

\label{lastpage}
\end{document}